\documentclass[12pt,a4paper]{article}

\usepackage{latexsym}
\usepackage{amsfonts}
\usepackage{amssymb}
\usepackage{amsmath}
\usepackage{graphicx}
\usepackage[dvips]{psfrag}
\usepackage{vmargin}
\usepackage{cite}
\usepackage{theorem}

\setpapersize{A4}
\setmargins{25mm}{20mm}{160mm}{220mm}{12pt}{20pt}{12pt}{36pt}

\newcommand\cmp[3]  {{\it Comm.\ Math.\ Phys.\ }{\bf #1} (#2) #3}

\newcommand\jmp[3]  {{\it J.\ Math.\ Phys.\ }{\bf #1} (#2) #3}

\newcommand\jhep[3] {{\it J. High Energy Phys.\ }{\bf #1} (#2) #3}

\newcommand\npb[3]  {{\it Nucl.\ Phys.\ }{\bf B #1} (#2) #3}

\newcommand\pla[3]  {{\it Phys.\ Lett.\ }{\bf A #1} (#2) #3}
\newcommand\plb[3]  {{\it Phys.\ Lett.\ }{\bf B #1} (#2) #3}

\newcommand\prd[3]  {{\it Phys.\ Rev.\ }{\bf D #1} (#2) #3}

\newcommand\ptp[3]  {{\it Prog.\ Theor.\ Phys.\ }{\bf #1} (#2) #3}
\newcommand\rmp[3]  {{\it Rev.\ Mod.\ Phys.\ }{\bf #1} (#2) #3}

\newcommand{\hepth}[1]{{\tt hep-th/#1}}

\newcommand{\scsc}{\scriptscriptstyle}

\newcommand{\bz}{\bar{z}}

\newcommand{\alb}{\bar{\alpha}}
\newcommand{\cd}{c^{\dag}}
\newcommand{\bra}[2]{\left<#1,#2\right|}
\newcommand{\cket}[2]{\left|#1,#2\right>}

\newcommand{\nh}{\hat{n}}
\newcommand{\Nh}{\hat{N}}

\theoremstyle{break}
\newtheorem{theorem}{Theorem}[section]
\theoremstyle{break}
\newtheorem{lemma}{Lemma}[theorem]
\theoremstyle{break}
\newtheorem{remark}{Remark}[theorem]

\begin{document}

\begin{titlepage}

\begin{flushright}
 \hepth{0209139}
\end{flushright}

 \vspace*{10mm}

 \begin{center}

  {\LARGE\sffamily\bf Instanton Number
    of \\ 
  Noncommutative U(n) gauge theory}
  \vspace{25mm}\\

  {\sffamily\bf 
  Akifumi Sako\footnote{{\it E-mail} :
  \ttfamily sako@math.sci.hiroshima-u.ac.jp}\vspace{15mm}}\\
  {\it Graduate School of Science, Hiroshima University,\\
  1-3-1 Kagamiyama, Higashi-Hiroshima 739-8526, Japan}\vspace{20mm}\\

 \end{center}


 \begin{center}
  {\bf ABSTRACT}
 \end{center}
We show that the
integral of the first Pontrjagin class is given by 
an integer
and it is identified with instanton number of the 
$U(n)$ gauge theory on noncommutative ${\bf R^4}$.
Here the dimension of the 
vector space $V$ that appear
in the ADHM construction is called  Instanton number.
The calculation is done in operator formalism
and the first Pontrjagin class is defined by converge series.
The origin of the instanton number is investigated closely, too.
 

\end{titlepage}

\section{Introduction}

Recently, there has been much interest in noncommutative field theory
motivated by the string theory and we can see it in
 \cite{Connes2,Douglas1,Seiberg2,Douglas2} and so on.
These discoveries show us the nonperturbative 
analysis of noncommutative gauge theory is
very important for the string theory.
Especially, instantons play the most essential 
role in nonperturbative analysis.
In commutative theory, there is a well-known method to construct
instanton solution, which is given by Atiyah, Drinfeld, Hitchin and
Manin (ADHM)~\cite{Atiyah1,Corrigan1,Osborn1}.
There is the one-to-one
correspondence between the instanton solutions and the ADHM data.
On the other hand, in noncommutative spaces case, a pioneering work for
instantons was done by Nekrasov and Schwarz \cite{Nekrasov1}.
They discovered deformed ADHM method to construct instanton of
noncommutative gauge theory. This deformation corresponds to the resolution of instanton 
moduli space \cite{Nakajima2}.
After that, many progress are reported about noncommutative 
instanton. For example, multi instanton solution of 
noncommutative U(1) gauge theory is given in \cite{Ishikawa1},
where the solution was discovered by the deformation of the commutative case
solution \cite{Braden1}.
Recently, Atiyah-Singer Index theorem for the noncommutative theory is discussed in 
\cite{Kim3}, that is related deeply with the instanton number. 
Other important progress are in 
\cite{Furuuchi1,Gross2,Nekrasov2,Nekrasov3,Lee1,Chu1,Kim1,Kim2,
Correa1,Lechtenfeld1,Kraus1,Kiritsis}.\\

Instanton is classified by topological charge that is 
called instanton number.
In commutative case, instanton number is
given by the integral of the 1st. Pontrjagin class 
and it is equivalent to ${\rm dim}~V$ in the ADHM construction
(see the next section).
However in noncommutative gauge theory, many problems had been left for the defining
 the integer-valued Pontrjagin class, and we had never found the proof 
to show that
the instanton number of $U(n)$ gauge theory is identified with the integral of the Pontrjagin class.
Here we define ``instanton number'' by the
${\rm dim}~V$, where 
$V$ is a complex vector space i.e.$V={\bf C}^k$ and it
appears in
ADHM construction (see section 2). 
(Thinking over the correspondence with the commutative case,
we had better use the name ``instanton number"for the 
integral of the first Pontrjagin 
class.
But the noncommutative instanton is discovered with ADHM construction
and the ``instanton number" is used as the dimension
of vector space $V$ in ADHM construction after \cite{Nekrasov1}. 
So, we are in accordance with this convention.)
For the noncommutative $U(1)$ theory 
and some special $U(n)$ cases that are essentially same as $U(1)$ case,
 Furuuchi explained the origin of the instanton number 
and identity between the instanton number and the Pontrjagin class 
\cite{Furuuchi1, Furuuchi2,Furuuchi4}.
Its strict proof was given in \cite{Ishikawa2}. 
These cases are easy to calculate instanton charge because we can put 
$J=0$ , where $J$ is  $Hom(V,~{\bf C^n})$ in ADHM data.
But we cannot adapt the condition $J=0$ to general $U(n)$ case. 
This had been the difficulty for proof for $U(n)$.\\

The aim of this article is to give the proof of identification
between the integral of the first Pontrjagin
class and ${\rm dim}~V$ for noncommutative $U(n)$ theory.
As same as \cite{Ishikawa2}, the integer-valued Pontrjagin class
is defined by the converge series.
The calculation is carried out by different way from \cite{Ishikawa2}
because we can not use $J=0$ condition.
As a result of the different way of the calculation, 
we can easily see the direct relation between ${\rm dim}~V$
and integral of the first Pontrjagin class and 
we can understand noncommutativity 
play a very important role there.
To clarify the distinction, we call the integral 
of the first Pontrjagin class ``instanton charge" 
and call $\dim$ of vector space $V$ ``instanton number".
\\

The organization of this paper is as follows.
In section 2, we review the noncommutative instanton and prepare
tools of instanton charge calculus.
In section 3, some lemmas are given to study 
the origin of the instanton number,
and some of them are necessary to calculate the instanton charge.
In section 4, we prove that the integral of the first Pontrjagin class is 
equal to ${\rm dim}~V$.
In section 5, we summarize this article.

\section{Noncommutative $U(n)$ Instantons}
We introduce the noncommutative field theory
used in this article and review the
noncommutative instantons, in this section.
\subsection{Noncommutative $\bf{R}^4$ and the Fock space representation}
Let us consider Euclidean noncommutative $\bf{R}^4$. Its coordinate
operators $x^{\mu}\;(\mu=1,2,3,4)$ on the noncommutative
manifold satisfy the following commutation relations
\begin{equation}
 [x^{\mu},x^{\nu}]=i\theta^{\mu\nu}, \label{EQ:xx-commu}
\end{equation}
where $\theta^{\mu\nu}$ is an antisymmetric real constant matrix,
whose elements are called noncommutative parameters.
We can always bring $\theta^{\mu\nu}$ to the skew-diagonal form
\begin{equation}
 \theta^{\mu\nu}=
  \begin{pmatrix}
   0            & \theta^{12} & 0            & 0 \\
   -\theta^{12} & 0           & 0            & 0 \\
   0            & 0           & 0            & \theta^{34} \\
   0            & 0           & -\theta^{34} & 0
  \end{pmatrix} \label{parameter}
\end{equation}
by space rotation.
We restrict the noncommutativity of the space to the
self-dual case of $\theta^{12}=\theta^{34}=-\zeta\ $.
In this article, we perform the calculation with this condition,
but $|\theta^{12}|=|\theta^{34}|$ is not essential.
This is 
only for convenience. On the other hand,
relative sign of the noncommutative parameter ( $\theta^{12}\theta^{34}>0$ ) is
significant and it is discussed in the section 5.
Here we introduce complex coordinates
\begin{equation}
 z_1=\frac{1}{\sqrt{2}}(x^1+ix^2),\;\; z_2=\frac{1}{\sqrt{2}}(x^3+ix^4),
  \label{EQ:z-coordinates}
\end{equation}
then the commutation relations (\ref{EQ:xx-commu}) become
\begin{equation}
 [z_1,\bz_1]=[z_2,\bz_2]=-\zeta,\;\;\mbox{others are zero}.
  \label{EQ:z-commu}
\end{equation}
For using the usual operator representation, we define creation and
annihilation operators by
\begin{equation}
 \cd_{\alpha}=\frac{z_{\alpha}}{\sqrt{\zeta}},\;\;
  c_{\alpha}=\frac{\bz_{\alpha}}{\sqrt{\zeta}},\;\;
  [c_{\alpha},\cd_{\alpha}]=1\;\;\;\;(\alpha=1,2).\label{EQ:c-a-operator}
\end{equation}
The Fock space $\cal H$ on which the creation and annihilation
operators (\ref{EQ:c-a-operator}) act is spanned by the Fock state
\begin{equation}
 \left|n_1,n_2\right>
  =\frac{(\cd_1)^{n_1}(\cd_2)^{n_2}}{\sqrt{n_1!n_2!}}\left|0,0\right>,
\end{equation}
with
\begin{eqnarray}
 c_1\cket{n_1}{n_2}=\sqrt{n_1}\cket{n_1-1}{n_2},\;\;&&
  \cd_1\cket{n_1}{n_2}=\sqrt{n_1+1}\cket{n_1+1}{n_2},\nonumber\\
 c_2\cket{n_1}{n_2}=\sqrt{n_2}\cket{n_1}{n_2-1},\;\;&&
  \cd_2\cket{n_1}{n_2}=\sqrt{n_2+1}\cket{n_1}{n_2+1},
\end{eqnarray}
where $n_1$ and $n_2$ are the occupation number.
The number operators are also defined by
\begin{equation}
 \nh_{\alpha}=\cd_{\alpha}c_{\alpha},\;\;\Nh=\nh_1+\nh_2,
\end{equation}
which act on the Fock states as
\begin{equation}
 \nh_{\alpha}\cket{n_1}{n_2}=n_{\alpha}\cket{n_1}{n_2},\;\;
  \Nh\cket{n_1}{n_2}=(n_1+n_2)\cket{n_1}{n_2}.
\end{equation}
In the operator representation, derivatives of a function
$f(z_1,\bz_1,z_2,\bz_2)$ are defined by
\begin{eqnarray}
 \partial_{\alpha}f(z)=[\hat{\partial}_{\alpha},f(z)],\hspace{5mm}
  \partial_{\bar{\alpha}}f(z)=[\hat{\partial}_{\bar{\alpha}},f(z)],
\end{eqnarray}
where $\hat{\partial}_{\alpha}=\bar{z}_{\alpha}/\zeta$,
$\hat{\partial}_{\bar{\alpha}}=-z_{\alpha}/\zeta$ which satisfy
\begin{equation}
 [\hat{\partial}_{\alpha},\hat{\partial}_{\bar{\alpha}}]=-\frac{1}{\zeta}.
\end{equation}
The integral on noncommutative $\bf{R}^4$ is defined by the standard
trace in the operator representation,
\begin{equation}
 \int d^4x=\int d^4z=(2\pi\zeta)^2\mbox{Tr}_{\cal H}.
\end{equation}
Note that $\mbox{Tr}_{\cal H}$ represents the trace over the Fock
space whereas the trace over the gauge group is denoted by
$\mbox{tr}_{U(n)}$.

\subsection{Noncommutative gauge theory and instantons}
Let us  consider the $U(n)$ Yang-Mills theory on noncommutative $\bf R^4$.
Let $M$ be a projective module over the algebra that is generated by
the $x_{\mu}$ satisfying (\ref{EQ:xx-commu}).

In the noncommutative space, the Yang-Mills connection is defined by
\begin{equation}
 \hat{\nabla}_{\mu}\psi=-\psi\hat{\partial}_{\mu}+\hat{D}_{\mu}\psi,
\end{equation}
where $\psi$ is a matter field in fundamental representation type 
and $D_{\mu} \in End(M)$ are anti-hermitian gauge
fields \cite{Nekrasov2}\cite{Nekrasov3}\cite{Gross2}.
The relation between $D_{\mu}$ and usual gauge connection $A_{\mu}$ is 
$D_{\mu}=-i\theta_{\mu \nu} x^{\nu} + A_{\mu}$, where $\theta_{\mu \nu}$
is an inverse matrix of $\theta^{\mu \nu}$ in (\ref{EQ:xx-commu}).
Then the Yang-Mills curvature of the connection $\nabla_{\mu}$ is
\begin{equation}
 F_{\mu\nu}=[\hat\nabla_{\mu},\hat\nabla_{\nu}]
  =-i\theta_{\mu\nu}+[{\hat D}_{\mu},{\hat D}_{\nu}].\label{EQ:F-x}
\end{equation}
In our notation of the complex coordinates (\ref{EQ:z-coordinates}) and
(\ref{EQ:z-commu}), the curvature (\ref{EQ:F-x}) is
\begin{equation}
 F_{\alpha\alb}=\frac{1}{\zeta}+[{\hat D}_{\alpha},{\hat D}_{\alb}],
  \hspace{5mm}
  F_{\alpha\bar{\beta}}=[{\hat D}_{\alpha},{\hat D}_{\bar{\beta}}]
  \hspace{10mm}(\alpha\not=\beta).  \label{curv}
\end{equation}
The Yang-Mills action is given by
\begin{equation}
 S=-\frac{1}{g^2}{\rm Tr}_{\cal H} \mbox{tr}_{U(n)}F\wedge
*F,\label{EQ:Y-M-action}
\end{equation}
where we denote ${\rm tr}_{U(n)}$ as a trace for the gauge group U(n),
$g$ is the Yang-Mills coupling and $*$ is Hodge-star.

Then the equation of motion is
\begin{equation}
 [\nabla_{\mu},F_{\mu\nu}]=0.\label{EQ:Y-M-eom}
\end{equation}
(Anti-)instanton solutions are special solutions of (\ref{EQ:Y-M-eom})
which satisfy the (anti-)self-duality ((A)SD) condition
\begin{equation}
 F=\pm*F.\label{EQ:solution-original}
\end{equation}
These conditions are rewritten in the complex coordinates as
\begin{eqnarray}
 F_{1\bar{1}}=&+&F_{2\bar{2}},\;\;F_{1\bar{2}}=F_{\bar{1}2}=0\;\;\;\;
  \mbox{(self-dual)},\label{EQ:SD}\\
 F_{1\bar{1}}=&-&F_{2\bar{2}},\;\;F_{12}=F_{\bar{1}\bar{2}}=0\;\;\;\;
  \mbox{(anti-self-dual)}.\label{EQ:ASD}
\end{eqnarray}
In the commutative spaces, solutions of Eq.(\ref{EQ:solution-original})
are classified by the topological charge
(integral of the first Pontrjagin class)
\begin{equation}
 Q=-\frac{1}{8\pi^2}\int\mbox{tr}_{U(n)}F\wedge F,\label{EQ:Q}
\end{equation}
which is always integer and called instanton number $k$.
However, in the noncommutative spaces above statement is unclear.
We discuss this issue in this article by using the operator
representation of (\ref{EQ:Q})
\begin{equation}
 Q=\begin{cases}
    \zeta^2\mbox{Tr}_{\cal H}\,\mbox{tr}_{U(n)}
    (F_{1\bar{1}}F_{2\bar{2}}-F_{12}F_{\bar{1}\bar{2}})\;\;\;\;&
    \mbox{(self-dual)}\\
    \zeta^2\mbox{Tr}_{\cal H}\,\mbox{tr}_{U(n)}
    (F_{1\bar{1}}F_{2\bar{2}}-F_{1\bar{2}}F_{2\bar{1}})\;\;\;\;&
    \mbox{(anti-self-dual)}.
    \end{cases}\label{EQ:Q-op-N}
\end{equation}
Note that this definition of instanton charge is naive one.
For determination of
the instanton charge, trace operation have to 
be defined by the summation over 
finite domain of the Fock space and the instanton charge
have to be defined by a converge series. These conditions are discussed
in the following sections. 

In this article, all studies and discussions are done
for ASD case, but they are adapted for SD case. 
\subsection{Noncommutative $U(n)$ instantons}
In the ordinary commutative spaces, there is a well-known way to find
ASD configurations of the gauge fields.
It is ADHM construction which is proposed by Atiyah, Drinfeld, Hitchin
and Manin \cite{Atiyah1}.
Nekrasov and Schwarz first extended this method to noncommutative
cases \cite{Nekrasov1}. 
 Here we review briefly the ADHM construction of $U(n)$
instantons\cite{Nekrasov2}\cite{Nekrasov3}.

The first step of ADHM construction on noncommutative $R^4$ is looking
for 
$B_1, B_2 \in End({\bf C}^k)$, $I\in Hom({\bf C}^n,{\bf C}^k)$ and 
$J\in Hom({\bf C}^k,{\bf C}^n)$ which satisfy the deformed ADHM equations
\begin{eqnarray}
 &&[B_1,B_1^{\dag}]+[B_2,B_2^{\dag}]+II^{\dag}-J^{\dag}J=2\zeta,
  \label{EQ:ADHM1}\\
 &&[B_1,B_2]+IJ=0.\label{EQ:ADHM2}
\end{eqnarray}
We call this $k$ ``instanton number''. 
In the previous section and abstract, $V$ denotes the 
vector space ${\bf C}^k $.
Note that the right hand side of Eq.(\ref{EQ:ADHM1}) is caused by
the noncommutativity of space.
The set of $B_1,B_2,I$ and $J$ is called ADHM data.
Using this ADHM data, we define operator 
${\cal D}:{\bf C}^k\oplus {\bf C}^k\oplus {\bf C}^n
\rightarrow {\bf C}^k\oplus {\bf C}^k$ by
\begin{eqnarray}
{\cal D}&=& \left(
\begin{array}{c}
\tau \\
\sigma^\dagger
\end{array}
\right) \nonumber \\
\tau&=&( B_2-z_2, B_1-z_1, I) = 
( B_2-\sqrt{\zeta}c_2^{\dagger}, B_1-\sqrt{\zeta}c_1^{\dagger}, I)\nonumber \\
\sigma^{\dagger} &=&
(-B^{\dagger}_1+\bar{z}_1, B^{\dagger}_2-\bar{z}_2, J^{\dagger})
=(-B^{\dagger}_1+\sqrt{\zeta}c_1, B^{\dagger}_2-\sqrt{\zeta}c_2, J^{\dagger}).
\end{eqnarray}
ADHM eqs. (\ref{EQ:ADHM1}) and (\ref{EQ:ADHM2}) are replaced by 
\begin{eqnarray}
\tau \tau^{\dagger} = \sigma^{\dagger} \sigma \equiv \square , 
\ \ \tau \sigma =0.  \label{ts=0}
\end{eqnarray}
The following fact about this $\square$ is known.
\begin{lemma}[Absence of zero-mode of $\square$]
There is no zero-mode
of $\square$. That is, for 
${}^{\forall} \lvert v \rangle \in {\cal H}\otimes{\bf C}^k$,
\begin{eqnarray}
\square \lvert v \rangle  = 0 \ \to \ \lvert v \rangle =0.
\end{eqnarray}
\label{lemma1}
\end{lemma}
Proof of this lemma is given by Furuuchi in the Appendix A of \cite{Furuuchi1}
 or by Nekrasov in the section 4.2 of \cite{Nekrasov2}.\\
 
Let $\Psi$ be the solutions of the equations:
\begin{eqnarray}
{\cal D} \Psi^a =0 \ \ (a=1,\cdots ,n) &&, \ \ \ \Psi^a: 
{\bf C}^n \rightarrow {\bf C}^k\oplus {\bf C}^k\oplus {\bf C}^n, \\
\Psi^{\dagger a} \Psi^b &=&\delta^{ab}  \ . \label{DP}
\end{eqnarray}

Then we can construct the $U(n)$ $k$ instanton(ASD) connection 
$D_\mu$ in (\ref{EQ:F-x}) as 
\begin{eqnarray}
D_\mu = i\Psi^{\dagger} \theta_{\mu \nu} x^{\nu} \Psi .
\end{eqnarray}
With the complex coordinate $z_{\alpha}$, they are rewritten as
\begin{eqnarray}
D_{\alpha}= \frac{1}{\zeta}\Psi^{\dagger} \bar{z_{\alpha}} \Psi
\ \ \ D_{\bar{\alpha}}= -\frac{1}{\zeta}\Psi^{\dagger} {z_{\alpha}} \Psi.
\end{eqnarray}

Let us enumerate important identities here.
They are obeyed from (\ref{DP}),(\ref{EQ:ADHM1}),
(\ref{EQ:ADHM2}) and (\ref{ts=0}). At first,
\begin{eqnarray}
\Psi \Psi^{\dagger} =
1-{\cal D}^{\dagger}\frac{1}{{\cal D D}^{\dagger}} {\cal D}.
\label{pp}
\end{eqnarray}
This is the most useful identity in this paper.
$\Psi \Psi^{\dagger}$ and 
${\cal D}^{\dagger}\frac{1}{{\cal D D}^{\dagger}} {\cal D}$ 
are projectors. 
$\Psi \Psi^{\dagger}$ projects out the eigenvectors of 
${\cal D}^{\dagger}\frac{1}{{\cal D D}^{\dagger}} {\cal D}$.
In other words, there are $\Psi \Psi^{\dagger}$ zero modes, 
which are given by 
${\cal D}^{\dagger}\frac{1}{{\cal D D}^{\dagger}} {\cal D}$  eigen state.
In section 3, we investigate them closely.
This operators are expressed with $\tau$ and $\sigma$ as
\begin{eqnarray}
{\cal D D}^{\dagger}= \left( 
\begin{array}{cc}
\square & 0 \\
0 & \square
\end{array}
\right), \ \ \ 
{\cal D}^{\dagger}\frac{1}{{\cal D D}^{\dagger}} {\cal D}
= \tau^{\dagger} \frac{1}{\square} \tau + 
\sigma \frac{1}{\square} \sigma^{\dagger}.
\label{ddd}
\end{eqnarray}
These relations are going to be used in the calculation 
for the instanton charge several times.

\section{Zero modes of $\Psi \Psi^{\dagger}$ and its nature}
\label{SEC:elongated}
In this section, we prepare to calculate the instanton charge 
(integral of the Pontrjagin class) and analyze the 
source of the instanton charge.
It is expected that the instanton charge is integer and
equal to instanton number as same as commutative case.
The aim of this article is to give the proof of this statement.
It is given in the section 4, but the method used in the proof 
is differ from the one used for $U(1)$ case.
Therefore, the relation between $U(1)$ and $U(n)$ about 
the origin of the instanton charge is not understood by the 
calculus in the section 4.
To supplement explanation for it, we study the origin of the 
instanton charge by the similar way of the $U(1)$ case \cite{Ishikawa2}.

\subsection{Zero-mode of $\Psi \Psi^{\dagger}$}
It is convenience to give some lemmas that describe the detail of the 
zero-mode of $\Psi \Psi^{\dagger}$.\\

\begin{lemma}[Zero-mode of $\Psi \Psi^{\dagger}$]
\label{lemma:zero-mode}
Suppose that $\Psi$ and $\Psi^{\dagger}$ are given in the previous 
section. 
The vector $ \lvert v \rangle  \in 
({\bf C}^k \oplus {\bf C}^k \oplus {\bf C}^n) \otimes {\cal H}$
satisfying
\begin{eqnarray}
 \Psi \Psi^{\dagger} \lvert v \rangle  = 
  \langle v \rvert \Psi \Psi^{\dagger} =0 , \ 
  \ \lvert v \rangle \neq 0
\end{eqnarray}
is said to be a zero mode of $\Psi \Psi^{\dagger}$.
Then the zero-modes are given by following three types:
\begin{eqnarray}
\lvert v_1 \rangle =\left(
\begin{array}{c}
      (-B_1 + \sqrt{\zeta}c_1^{\dagger}) \lvert u \rangle \\
       ( B_2 - \sqrt{\zeta}c_2^{\dagger} )\lvert u \rangle \\
        J \lvert u \rangle 
\end{array}
\right), \ \ 
\lvert v_2 \rangle =\left(
\begin{array}{c}
      (B^{\dagger}_2 - \sqrt{\zeta}c_2) \lvert u' \rangle \\
       ( B^{\dagger}_1 - \sqrt{\zeta}c_1 )\lvert u' \rangle \\
        I^{\dagger} \lvert u' \rangle
\end{array}
 \right) \\
\lvert v_0 \rangle =\left( 
\begin{array}{c}
       (\exp \frac{1}{\sqrt{\zeta}} \sum_{\alpha} B^{\dagger}_{\alpha} 
        c^{\dagger}_{\alpha}) \lvert 0 , 0\rangle v_{0}^i \\
       (\exp \frac{1}{\sqrt{\zeta}} \sum_{\alpha} B^{\dagger}_{\alpha} 
        c^{\dagger}_{\alpha}) \lvert 0 , 0 \rangle v_{0}^i\\
         0
\end{array}
\right).
\end{eqnarray}
Here $\lvert u \rangle$ ($\lvert u' \rangle$) is some element
of ${\bf C}^k \otimes {\cal H}$ 
(i.e. $\lvert u \rangle$ is expressed with the 
coefficients $u^{nm}_i \in {\bf C}$ as
$\lvert u \rangle = \sum_i \sum_{n,m} u^{nm}_i \lvert n,m \rangle e_i$ ,
where $e_i$ is a base of $k$-dim vector space  ).
$v_0^i$ is a element of $k$-dim vector. 
\label{lemma2}
\end{lemma}

{\bf Proof.}  From eq.(\ref{pp}), 
$\Psi \Psi^{\dagger}$ have a zero-mode as an eigenvector
of the operator ${\cal D}^{\dagger}\frac{1}{{\cal D D}^{\dagger}} {\cal D}$.
(In the following we call eigenvector
of the operator ${\cal D}^{\dagger}\frac{1}{{\cal D D}^{\dagger}} {\cal D}$
``${\cal D}$ eigenvectors". ``$\tau$ eigenvector" is used similarly.)
By the Eqs.(\ref{ts=0}), ${\cal D}$ eigen vectors are classified as
eigen vectors of $\tau$ or $\sigma^{\dagger}$.
The equation $\tau \sigma=0$ show that the 
$ \lvert v_1 \rangle$ ($ \lvert v_2 \rangle$) 
is in $\ker \tau$ ($\ker \sigma^{\dagger}$).
\begin{eqnarray}
\tau \lvert v_1 \rangle = \tau \sigma  \lvert u \rangle=0
, \ \sigma^{\dagger} \lvert v_2 \rangle =
\sigma^{\dagger} \tau^{\dagger}  \lvert u' \rangle=0.
\end{eqnarray}
If $\lvert u \rangle$ and $\lvert u' \rangle$ are not zero
vectors, as a result of the lemma\ref{lemma1},
\begin{eqnarray}
\sigma^{\dagger} \lvert v_1 \rangle = 
\sigma^{\dagger} \sigma  \lvert u \rangle 
=\square \lvert u \rangle \neq 0
&,&  \ \tau \lvert v_2 \rangle =
 \tau \tau^{\dagger}  \lvert u' \rangle 
 =\square \lvert u' \rangle
 \neq 0. \nonumber \\
{\cal D}^{\dagger}\frac{1}{{\cal D D}^{\dagger}} {\cal D}
\lvert v_1 \rangle = \sigma \lvert u \rangle = \lvert v_1 \rangle
&,& \ \ 
{\cal D}^{\dagger}\frac{1}{{\cal D D}^{\dagger}} {\cal D}
\lvert v_2 \rangle = \lvert v_2 \rangle.
\end{eqnarray}
If we chose $k$ independent vectors  
$\sum_{nm} u^{nm}_i \lvert n,m \rangle$ ($i=(1,\cdots ,k)$)
and orthonormalize them, we get ${\cal D}$ $k$-dim  eigen vector 
space that is zero-mode space of $\Psi \Psi^{\dagger}$.
$\sum_{nm} {u'}^{nm}_i \lvert n,m \rangle$ is similar to 
$\sum_{nm} u^{nm}_i \lvert n,m \rangle$ and we can 
construct other $k$-dim zero-mode space.\\
The remnant of this proof is for $\lvert v_0 \rangle$.
From commutation relations, 
\begin{eqnarray}
\sigma^{\dagger} \lvert v_0 \rangle =0,
\end{eqnarray}
and it shows that  $\lvert v_0 \rangle$ is belong to the eigen vectors 
of $\tau$. From the following relation, it is shown that
$\lvert v_0 \rangle$ is independent
from the  $\lvert v_2 \rangle$,
\begin{eqnarray}
\langle v_0 | v_2 \rangle =0.
\end{eqnarray}
After we orthonormalize them, it is concluded that 
they are eigen vectors
of ${\cal D}$ and zero-modes of $\Psi \Psi^{\dagger}$.
\begin{flushright}
$\square$
\end{flushright}
We are going to know the zero-mode $\lvert v_0 \rangle$ corresponds
to the source of the instanton charge and this zero-mode plays
a essential role when we introduce cut-off (boundary) for
the Fock space.
This lemma is not directly used in the instanton charge computation in section 4.
But this is important when we understand the origin of the instanton charge
by the parallel way of $U(1)$ instanton case.


\subsection{Boundary}
In the \cite{Ishikawa1,Ishikawa2}, 
 we introduced the Fock space cut-off $N$.
The cut off make the instanton charge be a converge series
and be a meaningful in these papers.
So we introduce cut-off, again.
But some complex rules of cut-off boundary
or trace operation  occur
as a result of the zero-mode $\lvert v_0 \rangle$.
At first, let us study the rules.

\begin{figure}[t]
  \begin{center}
   \begin{minipage}{70mm}
    \begin{center}
     \psfragscanon
     \psfrag{n1}[][][1.5]{$n_1$}
     \psfrag{n2}[][][1.5]{$n_2$}
     \psfrag{0} [][][1.5]{$0$}
     \psfrag{p1}[][][1.5]{$k-1$}
     \psfrag{p2}[][][1.5]{1}
     \psfrag{c1}[][][1.5]{$N$}
     \psfrag{c2}[][][1.5]{$N$}
     \psfrag{P} [][][1.5]{$P$}
     \psfrag{R} [][][2.0]{$\underbar{\it D}$}
     \scalebox{0.5}{\includegraphics{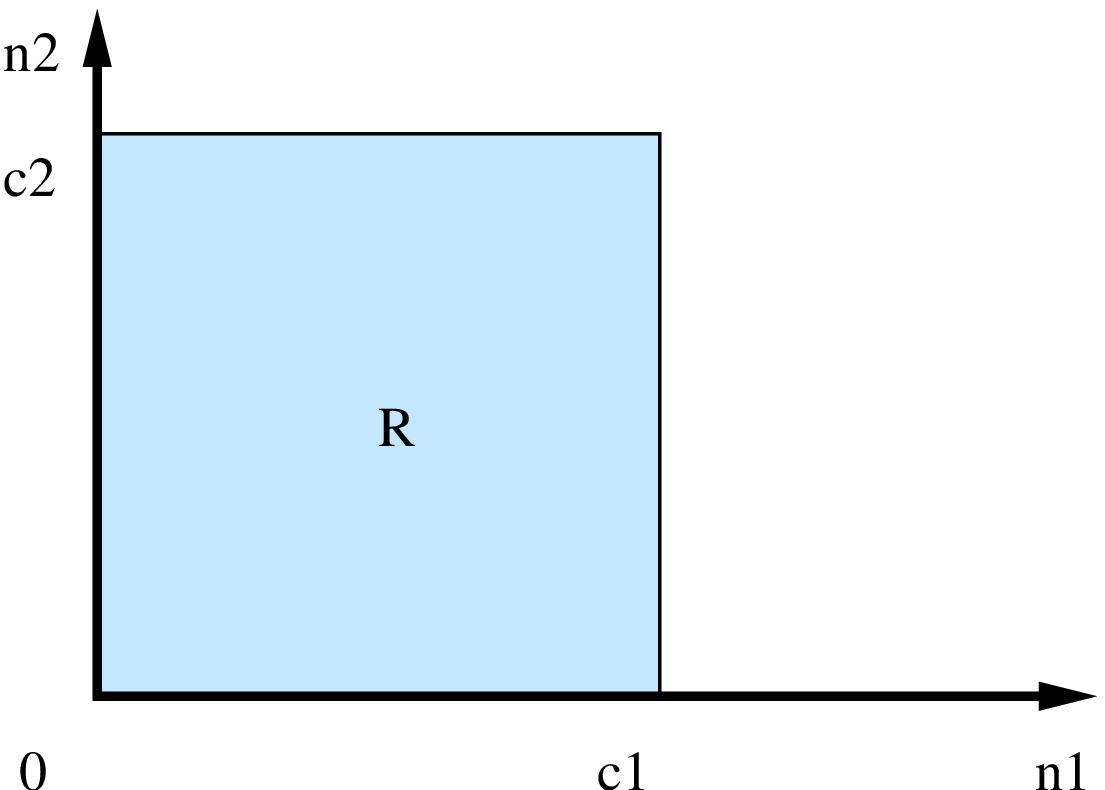}}
    \end{center}
    \caption{Region of $\mbox{Tr}_{\cal H}$ operation;
    This represents the region of summation of $\mbox{Tr}_{\cal H}$
    operation.
    We set the cut-off of the Fock space in the $n_1$ and $n_2$ direction at
$N$ in \cite{Ishikawa2}.}
    \label{FIG:region0}
   \end{minipage}
   \hspace{5mm}
   \begin{minipage}{70mm}
    \begin{center}
     \psfragscanon
     \psfrag{n1}[][][1.5]{$n_1$}
     \psfrag{n2}[][][1.5]{$n_2$}
     \psfrag{0} [][][1.5]{$0$}
     \psfrag{p1}[][][1.5]{$k-1$}
     \psfrag{p2}[][][1.5]{$0$}
     \psfrag{c1}[][][1.5]{$N$}
     \psfrag{c2}[][][1.5]{$N$}
     \psfrag{P} [][][1.5]{$ v_0$}
     \psfrag{R} [][][2.0]{$\bar{D}$}
     \psfrag{s1}[][][1.5]{$N+k$}
     \scalebox{0.5}{\includegraphics{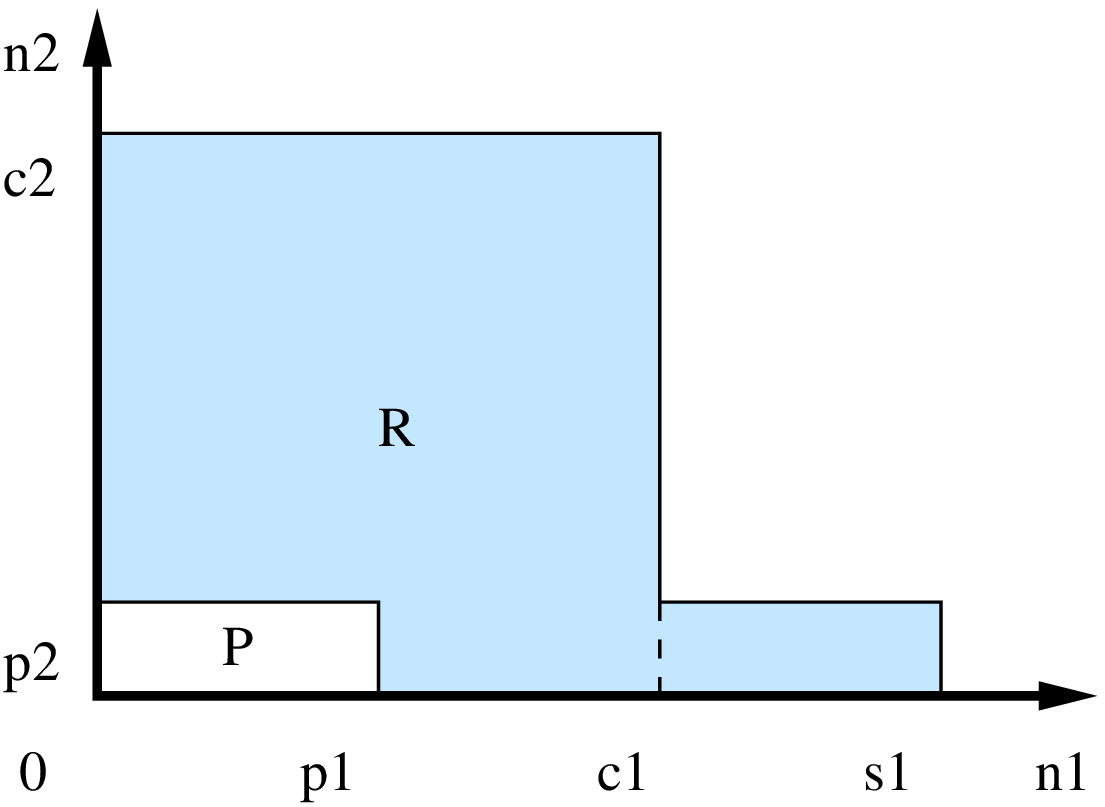}}
    \end{center}
    \caption{Region of the intermediate states;
    This represents the region of intermediate summation of the Fock space.
    Intermediate states appear inside of operators 
$\Psi^{\dagger}$ and $\Psi$  in the
U(n) ($S$ and $S^{\dag}$ in the U(1) ) case. \vspace{7mm}}
    \label{FIG:regionS}
   \end{minipage}
  \end{center}
\end{figure}

As the simplest example, we review the $U(1)$ case 
in \cite{Ishikawa1,Ishikawa2}.
In these paper, we set the cut-off in the $n_1$ and $n_2$ 
directions on $N$ ($N\gg k$)
(Fig.\ref{FIG:region0}),
here the ``cut-off'' means the trace operation cut-off, that is,
\begin{equation}
 {\rm Tr}_{\cal H}\left|_{\scsc [0,~N]}
     O\right.
 =
 \sum_{n_1=0}^{N}\sum_{n_2=0}^{N}
 \bra{n_1}{n_2}O\cket{n_1}{n_2} \;,
\end{equation}
where $O$ is an arbitrary operator.
Note that this cut off is only for the upper bound of summation of the
initial state and the final state. To the contrary, upper bound of
intermediate summation is frequently over $N$. 
In the $U(1)$ case (and $J=0$ case of $U(n)$), 
we can construct the instanton connection with
the  shift operator apparently. The shift operator cause the shift
of the boundary for the summation 
over intermediate states (see Fig.\ref{FIG:regionS}).
In \cite{Ishikawa1,Ishikawa2},
$N \times N$ square was chosen for the boundary of 
initial and final state. The instanton charge do not depend
on the shape of the Fock space region.
For example, 
$N_1 \times N_2$ rectangle or triangle that is defined by
$n_1 + n_2 = N$ are simple boundary for calculation.
In either case, we can get the same result of instanton charge.\\

\begin{figure}[t]
  \begin{center}
   \psfragscanon
   \psfrag{n1}[][][1.5]{$n_1$}
   \psfrag{n2}[][][1.5]{$n_2$}
   \psfrag{c1}[][][1.5]{$O(N)$}
   \psfrag{c2}[][][1.5]{$O(N)$}
   \psfrag{0} [][][1.5]{$0$}
   \psfrag{m} [][][1.5]{$m$}
   \psfrag{l} [][][1.5]{$l$}
   \psfrag{nm}[][][1.5]{$N_1(m)$}
   \psfrag{nl}[][][1.5]{$N_2(l)$}
   \scalebox{0.5}{\includegraphics{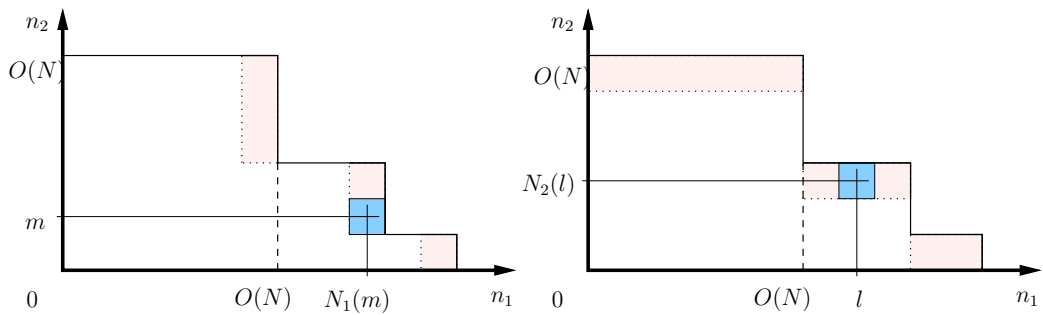}}
  \end{center}
 \caption{$N_1(m)$, $N_2(l)$: The set of the Fock states is surrounded 
 by the boundary whose length is order $N$.
 $N_1(N_2)$ depends on $n_1(n_2)$.}
 \label{FIG:N1m-N2l}
\end{figure}

To clarify the independence of the result from the shape of the boundary,
we do not make concrete shape of boundary, here.
Only the characteristic size of domain is introduced by $N$.
The region of the initial and final state of 
Fock space with the boundary is a set of states:
\begin{eqnarray}
\lvert n_1 , n_2 \rangle  ; 
n_1 = 0,\cdots , N_1(n_2) \ \ n_2 = 0,\cdots , N_2(n_1),
\end{eqnarray}
where $N_1(n_2)$ ($N_2(n_1)$) is a function of $n_2$ 
($n_1$) and  we suppose that length of the boundary is order $N \gg k$
i.e.
$N_1(n_2) \approx N_2(n_1) \approx N \gg k$ ( see Fig.\ref{FIG:N1m-N2l}). 
Using this cut-off (boundary), we define the instanton charge 
(integral of the first Pontrjagin class) by
\begin{eqnarray}
 Q&=&\lim_{N\rightarrow\infty}Q_N,\\
 Q_N&=&\zeta^2\sum_{n_1=0} \sum_{n_2=0}^{N_2(n_1)}
  \bra{n_1}{n_2}(F_{1\bar{1}}F_{2\bar{2}}-F_{1\bar{2}}F_{2\bar{1}})
  \cket{n_1}{n_2}. \label{Q}
\end{eqnarray}
One problem is left to determine (\ref{Q}), that is
to define the region of the Fock space of intermediate state.
As \cite{Ishikawa2} case, the region for summation of 
intermediate state is shifted. This phenomena is caused by the 
existence of the $\Psi \Psi^{\dagger}$ zero-mode.
Therefore we have to introduce the different region for intermediate
state.  (The state, which is sandwiched by $\Psi^{\dagger}$ from left
and $\Psi$ from right, is calld 
 ``intermediate state".)
For introducing it, the following remark is important.
\begin{remark}[Dimension of the zero-modes]
Consider the region of the Fock state as 
$\{ (n_1,n_2) | 0 < n_1 < \underbar{{\it N}}_1(n_2) , 
0 < n_2 < \underbar{{\it N}}_2(n_1) \}$.
(Under bar of $ \underbar{{\it N}}_i$ discriminate 
between boundary of initial states and one of intermediate state.)
If the boundary shape is convex,
then there is $\underbar{{\it N}} $ dimension Fock state vector
,where 
\begin{eqnarray}
\underbar{{\it N}} \equiv \sum_{n_2=0}  (\underbar{{\it N}}_1(n_2)+1) 
= \sum_{n_1=0}  (\underbar{{\it N}}_2(n_1)+1) .
\end{eqnarray}
Here $\sum_{n_2}  (\underbar{{\it N}}_1(n_2)+1)$ is the area of the 
region of the Fock state. 
($n_1=0$ line cause ``$+1$" in $(\underbar{{\it N}}_1(n_2)+1)$.)
Taking account of this fact, we can introduce dimension for subset of Hilbert space
(finite domain of the Fock space)
by counting each Fock state as an independent vector.
Then the dimension of the zero-modes 
$\left\{ \lvert v_1 \rangle \right\}$,
$\left\{ \lvert v_2 \rangle \right\}$ and 
$\left\{ \lvert v_0 \rangle \right\}$ are 
$k \times \underbar{{\it N}}$, $k \times \underbar{{\it N}}$ and $k$.
\label{dim}
\end{remark}

Note that above boundary is only for one component of ($2k+n$)-dim 
vector (remember that
$\Psi^a \in  {\bf C}^{2k+n} \otimes {\cal H}$).
In general, each component of the vector do not
have universal boundary.
So if the boundary is not universal, some index $i$ ($i \in \{1,\cdots , 2k+n \} $) 
have to be introduced to assign to each boundary i.e. 
$\underbar{{\it N}} \rightarrow \underbar{{\it N}}^i$.\\ 
From this remark, to make $\Psi$ be an isomorphism for finite domain Fock space,
the dimension of initial state
should be the same dimension of intermediate state.
(Note that $\Psi$ is a partial isometry, i.e. $\Psi \Psi^{\dagger} \neq 1$ for infinite dimensional Fock
space, but it is possible to be regarded as an isomorphism when we consider
 finite domain.) 
For some fixed $l$ and $p$, $\Psi^a_{l,p,n_1,n_2} \lvert n_1,n_2 \rangle$ is 
a expanded by $n \times \underbar{{\it N}}$ bases.
On the other hand, when left module Fock space is defined by the same domain 
as the right module, for some fixed $n$ and $m$, 
$ \langle l,p \rvert \Psi^a_{l,p,n,m}$ is expanded by 
 $n \times \underbar{{\it N}}-k$ bases because there are $k$ zero-modes
 $\langle v_0 \rvert$ that are removed from only intermediate state.
Therefore we have to set the cut-off of the intermediate state as
\begin{eqnarray}
\sum_{i=1}^{2k+n}\sum_{n_2=0} (\bar{N}^i_1(n_2)+1)
=\sum_{i=1}^{2k+n}\sum_{n_1=0} (\bar{N}^i_2(n_1)+1) =(2k+n) \times 
\underbar{{\it N}} +k, \label{inter}
\end{eqnarray}
where $\bar{N}^i_{\alpha}$ is  boundary of intermediate state for 
$i$-th component of $(2k+n)$-dim vector.
This boundary is naive extension of the $U(1)$ 
case (Fig.\ref{FIG:regionS}).\\

Generally, it is difficult to give a concrete expression of intermediate boundary  
from initial state boundary. 
But intermediate boundary is often more necessary than initial boundary
for instanton charge calculation in the next section. 
Therefore we define the trace operation of finite region not by initial state domain but intermediate domain, in the next section. 
It is tricky but very convenience definition.  \\

It is worth to mention 
the essence of the above finite domain of the trace operation.
It is to make $\Psi$ be an isometry for finite domain.
So the domain of the intermediate state is changed by the situation.
For example, consider two state
$\Psi^{\dagger} \cdots \langle inter |\Psi |initial \rangle$
and $\Psi^{\dagger} \cdots \langle inter' |c_1^{\dagger} \Psi |initial \rangle$.
Then the domain (set) of the intermediate state $  \langle inter' |$ is 
given by domain of $  \langle inter |$ shifted by $c_1^{\dagger}$.
Therefore region of ``intermediate state" is determined case by case. \\

To support the definition of the boundary and above discussions,
following Remark is important.
\begin{remark}[Asymptotic behavior of the $\lvert v_0 \rangle$]
\label{asymptotic}
If $N_1 + N_2 \equiv N$ is enough large, using some positive 
constant $c$ and $\epsilon <1$,
\begin{eqnarray}
\lvert \langle N_1 , N_2|\otimes 1 \ |v_0 \rangle \rvert
 < {c}{\epsilon^N}.
 \label{asympt}
\end{eqnarray}
\end{remark}
{\bf Proof.} Some constant $\beta$ exists, which satisfies 
\begin{eqnarray}
\lvert \langle N_1 , N_2|\otimes 1\ 
| v_0 \rangle \rvert &=&
\left|
 \left( 
 \begin{array}{c}
  \langle N_1 , N_2   |   (\exp \sum_{\alpha} B^{\dagger}_{\alpha} 
        c^{\dagger}_{\alpha}) | 0 , 0\rangle v_{0}^i \\
  \langle N_1 , N_2    | (\exp \sum_{\alpha} B^{\dagger}_{\alpha} 
        c^{\dagger}_{\alpha}) | 0 , 0 \rangle v_{0}^i\\
         0
 \end{array}
 \right) 
\right|
 \nonumber \\
&<& 
\left|
\left(
 \begin{array}{c}
  \frac{\beta^{N_1+N_2}}{\sqrt{N_1! N_2!}}\\
  \frac{\beta^{N_1+N_2}}{\sqrt{N_1! N_2!}}\\
         0
 \end{array}
\right)
\right| .
\end{eqnarray}
When $N_1+N_2 \equiv N$ is large enough and we assume $N_1 > N_2$. 
($N_2 > N_1 $ case is proved as same as following way.) 
Then 
\begin{eqnarray}
{}^\exists N'\ |\  \beta < {N'}^{\frac{1}{4}} <N' <\frac{N}{2}. 
\end{eqnarray}
Using this $N'$,
\begin{eqnarray}
\frac{\beta^{N_1+N_2}}{\sqrt{N_1! N_2!}}
&<& \frac{\beta^{N}}{\sqrt{N_1!}}
< \frac{\sqrt{N'}^{\frac{N}{2}}}{\sqrt{N_1!} }
<\frac{\sqrt{N'}^{N_1}}{\sqrt{N_1!} }
\nonumber \\
&=& \frac{\sqrt{N'}^{N'}}{\sqrt{N'!} }
 \sqrt{\frac{(N')^{N_1-N'}}{(N'+1)\cdots N_1}}
<\frac{\sqrt{N'}^{N'}}{\sqrt{N'!} }
  \left( \sqrt{\frac{N'}{(N'+1)}} \right)^{N_1-N'} < c' {\epsilon}^N ,
\end{eqnarray}
where $c'= \sqrt{\frac{(N'+1)^{N'}}{N'!}}$ and 
$ \epsilon = \left( \frac{N'}{N'+1} \right)^{\frac{1}{4}} $.
After replacing $ 2kc'$ by $c$,
we get eq.(\ref{asympt}).
\begin{flushright}
$\square $
\end{flushright}
From this remark, we can ignore the $v_0$ effect near the boundary for large
$N$ up to exponential damping factors, in calculations of section 4, and it is possible that 
 boundary for trace operation is
introduced with no obstruction caused by the zero mode $v_0$.

\subsection{ projector near boundary}
We will use concrete expression of $\Psi \Psi^{\dagger} $,
in section 4. So, we give it here.
From Eqs. (\ref{pp}) and (\ref{ddd}), $\Psi \Psi^{\dagger} $
 is written by using ${\cal D}$ and 
$1/\square $, where 
\begin{eqnarray}
\square =\sum_{\alpha=1,2}
(B_{\alpha} {B_\alpha}^{\dagger} + \zeta \hat{N} 
 -\sqrt{\zeta} \vartriangle +II^{\dagger}).
\end{eqnarray}
 $\hat{N}$ denotes total number operator $\hat{n}_1+\hat{n}_2$ and 
define $\vartriangle$ by
\begin{eqnarray}
\vartriangle = \sum_{\alpha=1,2}
c_{\alpha} B_{\alpha} + c^{\dagger}_{\alpha} B^{\dagger}_{\alpha}. 
\end{eqnarray}
For Fock states with large eigen value of $\hat{N}$,
we can get the concrete form of $\square^{-1}$ as
\begin{eqnarray}
\frac{1}{\square }=
\frac{1}{\zeta N} + 
\frac{1}{{\zeta}^{\frac{3}{2}}} \frac{1}{N} \vartriangle \frac{1}{N}+
\frac{1}{{\zeta}^{2}}\frac{1}{N}
(-\sum B_{\alpha} {B_\alpha}^{\dagger} -II^{\dagger}
\vartriangle \frac{1}{N} \vartriangle )\frac{1}{N}
+O(N^{-\frac{3}{2}}),
\label{sq_-1}
\end{eqnarray}
where $N$ is eigenvalue of $\hat{N}$.
Unless it make misunderstanding, we do not distinguish operators $\hat{n}_1$, 
$\hat{n}_2$ and $\hat{N}$ from their eigenvalue $n_1$ $n_2$ and $N$ in the following. 
To express $ \Psi \Psi^{\dagger} $ near the boundary
, we use following notation:
\begin{eqnarray} \label{ppd}
\Psi \Psi^{\dagger} 
&\rightarrow &
\left(
\begin{array}{ccc}
\Psi \Psi^{\dagger}_{11} & \Psi \Psi^{\dagger}_{12} & 
\Psi \Psi^{\dagger}_{13} \\   
\Psi \Psi^{\dagger}_{21} & \Psi \Psi^{\dagger}_{22} & 
\Psi \Psi^{\dagger}_{23} \\
\Psi \Psi^{\dagger}_{31} & \Psi \Psi^{\dagger}_{32} & 
\Psi \Psi^{\dagger}_{33} 
\end{array}
\right) + O(N^{-\frac{3}{2}}) \\
&=& \left(
\begin{array}{ccc}
O\!\left(\frac{1}{N} \right) & O\!\left(\frac{1}{N} \right) & 
O\!\left(\frac{1}{\sqrt{N}} \right)\\
 O\!\left(\frac{1}{N} \right) & O\!\left(\frac{1}{N} \right) & 
O\!\left(\frac{1}{\sqrt{N}} \right)\\
O\!\left(\frac{1}{\sqrt{N}} \right)& O\!\left(\frac{1}{\sqrt{N}}\right) & 1_{[n]}
\end{array}
\right) .
\end{eqnarray}
Note that $\Psi \Psi^{\dagger}_{11}, \Psi \Psi^{\dagger}_{12}
\Psi \Psi^{\dagger}_{21}$ and $ \Psi \Psi^{\dagger}_{22}$ are
$k \times k$ matrices.
$\Psi \Psi^{\dagger}_{31}$ and $\Psi \Psi^{\dagger}_{32}$
are
$n \times k$.
$\Psi \Psi^{\dagger}_{13}$ and $\Psi \Psi^{\dagger}_{23}$
are $k \times n$ matrices.
$\Psi \Psi^{\dagger}_{33}$ is a $n \times n$ matrix and $1_{[n]}$ 
is a $n \times n$  identity.
Using (\ref{sq_-1}) and (\ref{pp},\ref{ddd}), these are written as
\begin{eqnarray}
\Psi \Psi^{\dagger}_{11} &=& 
-\frac{1}{N}\left( 1-\frac{1}{\zeta} II^{\dagger} 
-\frac{1}{\zeta}[B_2 , B_2^{\dagger}] 
+\frac{n_1-n_2}{N} \right)   \label{ppp1}\\
\Psi \Psi^{\dagger}_{22} &=& 
-\frac{1}{N}\left( 1-\frac{1}{\zeta} II^{\dagger} 
-\frac{1}{\zeta}[B_1 , B_1^{\dagger}] 
+\frac{n_2-n_1}{N} \right)  \label{ppp2}\\
\Psi \Psi^{\dagger}_{33}&=&1-\frac{1}{\zeta N}(I^{\dagger}I+JJ^{\dagger})
\label{ppp3} \\
\Psi \Psi^{\dagger}_{21}&=&(\Psi \Psi^{\dagger}_{12})^{\dagger}
=-\frac{1}{\zeta N^2}(Ic_1 +J^{\dagger} c_2^{\dagger})
(-I^{\dagger}c_2^{\dagger} + J c_1 ) \label{ppp4}\\
\Psi \Psi^{\dagger}_{31}=(\Psi \Psi^{\dagger}_{13})^{\dagger}
&=& I^{\dagger} \frac{1}{\sqrt{\zeta} N} c_1
  -J\frac{1}{\sqrt{\zeta} N} c_2^{\dagger} \nonumber \\
&& -I^{\dagger}\frac{1}{\zeta N} B_2 
 +J\frac{1}{\zeta N} B_1^{\dagger} 
  +I^{\dagger}\frac{1}{\zeta N} \vartriangle \frac{1}{N}c_2^{\dagger} 
  -J \frac{1}{\zeta N} \vartriangle \frac{1}{N}c_1 \label{ppp5}\\
\Psi \Psi^{\dagger}_{32}=(\Psi \Psi^{\dagger}_{23})^{\dagger}
&=& I^{\dagger} \frac{1}{\sqrt{\zeta} N} c_1^{\dagger}
  +J\frac{1}{\sqrt{\zeta} N} c_2 \nonumber \\
&& -I^{\dagger}\frac{1}{\zeta N} B_1 
 -J\frac{1}{\zeta N} B_2^{\dagger} 
  +I^{\dagger}\frac{1}{\zeta N} \vartriangle \frac{1}{N}c_1^{\dagger} 
  -J \frac{1}{\zeta N} \vartriangle \frac{1}{N}c_2  .
\label{ppp6}
\end{eqnarray}

\subsection{Rough estimation of instanton charge}
We prove that instanton charge is equal to instanton number 
by direct calculation of integral of the first Pontrjagin class
in the next section.
In the calculation, we do not use the method that is used 
in \cite{Ishikawa2}. 
In \cite{Ishikawa2}, the origin of the instanton charge is 
clear that is zero-mode $v_0$. 
On the other hand, the method in the next section is simple
but unclear about the relation between the origin of the instanton charge
and the zero-mode $v_0$.
So, in this subsection, we do rough estimation of the 
instanton charge by the same method of \cite{Ishikawa2}
to complement understanding of the origin. \\

The aim of this subsection is not to give a strict proof but
to understand the origin of the 
instanton number by the zero-mode $v_0$, therefore the calculation is not strict.
In this subsection we ignore some terms appearing in the Pontrjagin class 
without explanation.
For example, we can not use the condition ${\cal D}\Psi=0$ on the boundary
of the finite Hilbert space in strict computation,
because non-zero terms appear in ${\cal D}\Psi$.
But we do not count their contribution in this subsection.
As we will see in the next section, their terms do not vanish
under large $N$ limit but they cancel out at last.\\

Using the Eq.(\ref{pp}) and the condition ${\cal D}\Psi=0$ with no-limitation 
(strict speaking, this is not correct as we will see in the next section), 
instanton charge (\ref{EQ:Q-op-N}) have following form:
\begin{eqnarray}
Tr_N 1 -Tr_N (\frac{1}{2} [\Psi^{\dagger} c_2^{\dagger} \Psi \ ,\ \Psi^{\dagger} c_2 \Psi]
+\frac{1}{2} [\Psi^{\dagger} c_1^{\dagger} \Psi \ ,\ \Psi^{\dagger} c_1 \Psi] ).
\label{ad}
\end{eqnarray}
Here, $ Tr_N $ denotes trace over some finite domain of Fock space
characterized by $N$ 
and it includes $tr_{U(n)}$ operation. 
Using the Stokes' like theorem in \cite{Ishikawa2}, 
trace over the boundary is obtained,
then 
$ Tr_N [\Psi^{\dagger} c_2^{\dagger} \Psi \ ,\ \Psi^{\dagger} c_2 \Psi] $ become
\begin{eqnarray}
\sum \Psi^{\dagger ai}_{n_1,m_1,l_1,p_1}\sqrt{p_1+1} 
(\Psi \Psi)^{\dagger ik}_{l_1,p_1+1,l_2,P_2}\sqrt{P_2}
\Psi_{l_2,P_2-1,n_1,m_1}^{ka}. \label{r1}
\end{eqnarray}
Here $\Psi_{l_1,p_1,n_2,m_2}^{ka}= 
\langle l_1,p_1 | \Psi^{ka} |n_2 , m_2 \rangle $.
$(l_i , p_i) \in \bar{D} $(intermediate state) , 
$(n_i , m_i) \in \underbar{{\it D}}$(initial state) and $(l_2 , P_2)$ is a state
on the boundary, (see Fig.\ref{FIG:region0}, Fig.\ref{FIG:regionS}, 
,Fig.\ref{FIG:barD} and section 4).
For enough large domain, the leading of the $(\Psi \Psi^{\dagger})_{l,p,l',p'}$
is equal to $1_{[n]} \delta_{l,l'} \delta_{p,p'}$ near the boundary, in other words
the gauge connection approaches to the pure gauge.
Then Eq.(\ref{r1}) is the same as (\ref{inter}):
\begin{eqnarray}
\sum_i \sum_{boundary} ({\bar{N}^i}_2(n_1)+1) = Tr_N 1 + k. \label{r3}
\end{eqnarray}
The same value is obtained from 
$Tr_N [\Psi^{\dagger} c_1^{\dagger} \Psi \ ,\ \Psi^{\dagger} c_1 \Psi]$ , too.
The first term of the right hand side of Eq.(\ref{r3}) 
and the first term in Eq.(\ref{ad}) cancel out. 
Note that the first term in Eq.(\ref{ad}) is 
from the constant curvature in (\ref{curv}). 
Finally the second term of Eq.(\ref{r3}) is understood as the source
of the instanton charge.
In other words, the origin of the instanton charge is $k$ in (\ref{inter}).
Recall that the reason for the $k$ is the zero-modes $v_o$ whose dimension is $k$. 
This rough estimation implies that the origin of the instanton number is
similar to the U(1) case \cite{Furuuchi1,Furuuchi2,Ishikawa2}.

\section{Explicit Calculation of the Instanton Charge}

In this section, we prove the following theorem.
\begin{theorem}[Instanton number]
\label{theorem1}
Consider U(n) gauge theory on noncommutative ${\bf R^4}$ 
whose commutation relations are given by self-dual relation.
( Especially we use Eqs.(\ref{EQ:z-commu}) for simplicity in the below proof.)
Then the integral of the first Pontrjagin class is possible to 
be defined by converge series and it is identified with
the dimension $k$ that is called ``instanton number" appearing in the ADHM construction in section 2.
\end{theorem}

\subsection{Rules of the calculation}
In this subsection, we compile the rules of calculation
that will be used in the next subsection.
In the previous section, we estimate the 
leading of the integral of the first Pontrjagin class
by using the Stokes' like theorem.
But it is not complete calculation because we have to 
estimate next leading contribution.
There are two types of calculation to get it.
First is to use only Stokes' like theorem as we did in 
$U(1)$ case \cite{Ishikawa2}.
The $U(1)$ case have  simple one kind of boundary to define
$Q_N$, as a result of $J=0$.
On the other hand, in this $U(n)$ case, the boundary of trace operation 
is complex, i.e. there is no universal boundary among 
each component, then we have to treat the summation over the 
boundary carefully.
So using the Stokes' like theorem
is tough and complex in $U(n)$ case.
Another way to evaluate the integral of the first Pontrjagin class 
is to introduce more concrete expression of the 
boundary and analyze the ADHM constraint 
on the boundary as we will do in the next subsection. 
This method have a advantage that the calculation is easier than 
using only Stokes' like theorem.
On the other hand, we do not use $v_0$ and the Lemma\ref{lemma:zero-mode} explicitly.
So the origin of the instanton number is understood not by $v_0$ but 
another way, which is going to be found by 
$tr (I^{\dagger} I -J J^{\dagger})/\zeta=2k$, at last.
As preparations for this calculation, we introduce concrete boundary
 and investigate its nature, here.\\

\begin{figure}[t]
  \begin{center}
   \scalebox{0.3}{\includegraphics{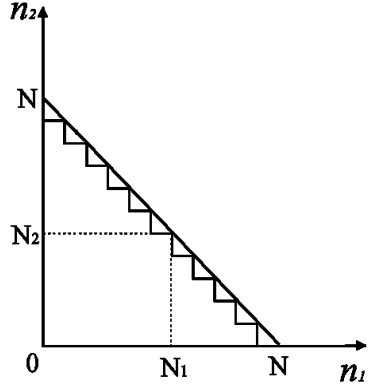}}
  \end{center}
 \caption{$\bar{D}$ : $N_1 + N_2=N$ determined the set of intermediate states
$\bar{D}$. It is used for the definition of the trace operation.}
 \label{FIG:barD}
\end{figure}

The aim is to make calculation be simple and easy, 
so we define the domain of trace operation
 by using following initial and final state.
\begin{eqnarray}
\underbar{{\it D}} \equiv \{  |initial\rangle \ | \ 
\langle n_1 , n_2 | 1_{[2k+n]} \Psi |initial\rangle\neq 0 \ , \ 
n_1 + n_2 \le N \} \label{domain} \\
\{  |final\rangle \ | \ 
\langle final | \Psi 1_{[2k+n]}|n_1 , n_2 \rangle\neq 0 \ , \ 
n_1 + n_2 \le N \}.
\end{eqnarray}
Here $1_{[2k+n]}$ is a $(2k+n)\times (2k+n)$ identity matrix
Let $\{ |initial\rangle \}$ be orthonormalized basis in the following.
As we saw in previous section, the Fock space dimension of $|initial\rangle$
and $|final \rangle$ is $-k+n(N+1)(N+2)/2$ for enough large $N$.
For convenience, we introduce the domain $\bar{D}$ (Fig.\ref{FIG:barD}):
\begin{eqnarray}
\bar{D} \equiv \{ |n_1 , n_2 \rangle | n_1 + n_2 \le N \}.
\end{eqnarray}

With these initial (or final) states, we define trace operation with boundary as
\begin{eqnarray}
Tr_N \Psi^{\dagger} O \Psi &\equiv & tr
\sum_{all \ \langle initial |} \langle initial | 
\Psi^{\dagger} O \Psi  |initial \rangle \label{trace} \\
=&tr& \!\!\!\!\!\!\!
\sum_{\scriptsize \begin{array}{c} n_1+n_2\le N \\
n_1'+n_2'\le N \end{array}} \sum_{all \ \langle initial |}
\langle n_1, n_2 | O | n_1' , n_2' \rangle 
\langle n_1, n_2 | \Psi |initial \rangle 
\langle initial | \Psi^{\dagger} | n_1' , n_2' \rangle,
\nonumber 
\end{eqnarray}
where $tr$ denotes trace of any other kind of indices without 
Fock space, that is the trace of $(2k+n)\times (2k+n)$ dimension matrices 
or gauge group U(n). 
Here, it is worth to emphasize 
the merit of definition (\ref{domain}) and (\ref{trace}).
In the computation of the Pontrjagin class, 
${\cal D}\Psi  |initial \rangle$ or $\langle final | \Psi^{\dagger}{\cal D}^{\dagger}$
often appear.
If there is no boundary, then the zero mode $\Psi $ satisfies
 ${\cal D}\Psi  =0$ by definition. However, we have boundary and the Fock space is finite,
then the condition is not satisfied on the boundary in general. 
Because the operator ${\cal D}$ includes creation and annihilation operators,
cancellation of each state is occurred by chain reaction.
On the boundary, the chain of cancellation is broken.
For example, if we choose (\ref{domain}) as the domain and 
$N_1+N_2 =N$ as boundary of intermediate state ,
following non-vanishing terms remain:
\begin{eqnarray}
\sum_{all \ |initial \rangle} {\cal D}\Psi  |initial \rangle &=&
\sum_{all \ |initial \rangle}\ \sum_{N_1+N_2=N} {\cal D}_b |N_1 , N_2 \rangle 
1_{[2k+n]} \langle N_1 , N_2 | \Psi  |initial \rangle
\nonumber \\
 = \sum_{all \ |initial \rangle}\ \sum_{N_1+N_2=N}&& \!\!\!\!\!\!\!\!\!\!\!\!
\left(
\begin{array}{ccc}
-\sqrt{\zeta} c_2^{\dagger} & -\sqrt{\zeta} c_1^{\dagger} & 0 \\
-B_1^{\dagger} & B_2^{\dagger} & J^{\dagger} 
\end{array}
\right) |N_1 , N_2 \rangle 1_{[2k+n]} \langle N_1 , N_2 | \Psi  |initial \rangle. \label{Db}
\end{eqnarray}
Here ${\cal D}_b$ denotes a operator whose
operands are boundary states of intermediate state 
(in this case, intermediate state belong to $\bar{D}$)
 and it 
is expressed as 
\begin{eqnarray}
\sum_{(n_1,n_2) \in domain}\!\!\!\!\!\!\!\!\!\!
 {\cal D} |n_1 , n_2 \rangle 
1_{[2k+n]} \langle n_1 , n_2| \Psi 
= \!\!\!\!\!\!\!\!\!\! 
\sum_{(N_1,N_2)\in boundary}\!\!\!\!\!\!\!\!\!\! {\cal D}_b |N_1 , N_2\rangle 
1_{[2k+n]} \langle N_1 , N_2| \Psi.
\end{eqnarray} 
Using (\ref{Db}), we can perform the instanton charge calculation
concretely and easily. This is the most important merit
to introduce the boundary  (\ref{domain}).\\

Note that the ${\cal D}_b$ do not have universal form for each intermediate
boundary state. 
i.e. ${\cal D} \Psi$ on boundary is not always expressed by Eq.(\ref{Db}).
As an  example, let us study the case of 
\begin{eqnarray}
\sum_{all \ |initial \rangle} 
{\cal D}\Psi \Psi^{\dagger} c_{\alpha} \Psi |initial \rangle .
\label{ex1}
\end{eqnarray}
If there is no boundary, ${\cal D}\Psi \Psi^{\dagger}$ is 
zero by the definition of the zero-mode $\Psi$.
But in (\ref{ex1}), there is boundary at
$\Psi |initial \rangle $ with the definition (\ref{domain}).
Then $\Psi \Psi^{\dagger} c_{\alpha} \Psi |initial \rangle $
is also expanded by finite number of the 
basis of the Fock space, 
and there is intermediate boundary.
In this case, intermediate state 
$| inter \rangle  $ is defined by complete system that can 
expand the space 
$\{ \Psi \Psi^{\dagger} c_{\alpha} \Psi |initial  \rangle  \} $, and
its $i$-component $| inter \ (i) \rangle $ satisfies
\begin{eqnarray}
{}^\exists  | initial \rangle &:& \sum_j
   \langle inter \ (i)|
( \Psi \Psi^{\dagger} c_{\alpha} \Psi )_{ij} |initial \ (j) \rangle 
\neq 0 .
\end{eqnarray}
We call set of the finite number of  intermediate states $\bar{I} $ i.e.
$ \bar{I}  = \{ | inter \rangle \} $ .
Then the boundary $|{\bf int.\!\!\! -b(i)} \rangle$ of the intermediate state is 
defined by 
\begin{eqnarray}
| {\bf int.\!\!\! -b(i)} \rangle &\equiv & b_{n_1,n_2,i}| n_{1}^{(i)}, n_{2}^{(i)}  \rangle
\\
\big( 
{}^\exists \langle inter \ (i) | &:&
\langle inter \ (i) | n_{1}^{(i)}, n_{2}^{(i)}\rangle \neq  0 \big),
\nonumber \\
&{\rm and }& \left\{ 
\begin{array}{c}
 \big( {}^\forall \langle inter \ (i) | : 
\langle inter \ (i) | n_{1}^{(i)}+1, n_{2}^{(i)}\rangle = 0 \big) \\
\ {\rm or }  \\
\big({}^\forall \langle inter \ (i) | : 
\langle inter \ (i) | n_{1}^{(i)}, n_{2}^{(i)} +1 \rangle = 0
\big) \end{array} 
\right\} .
\nonumber
\end{eqnarray}
For any $|initial \rangle $,
\begin{eqnarray}
{\cal D} \Psi \Psi^{\dagger} c_{\alpha} \Psi 
 |initial \rangle 
&=&  \sum_{{\bf int.\!\!\! -b}}{\cal D}_b | {\bf int.\!\!\! -b} \rangle 
 \langle {\bf int.\!\!\! -b} | 
\Psi \Psi^{\dagger} c_{\alpha} \Psi 
 |initial \rangle  \ . 
\end{eqnarray}
We have to pay attention to the point that 
this intermediate boundary is not possible 
to be expressed by the simple one condition like $N_1+N_2=N$
in (\ref{Db}),
 since $\Psi \Psi^{\dagger} c_{\alpha}$ make the boundary 
 be complex.
So we can not express the explicit expression 
of the ${\cal D}_b | {\bf int.\!\!\! -b} \rangle $.
Fortunately, we do not need its explicit expression in the calculation 
of the instanton charge in the next subsection, since 
the leading terms contributing to the instanton charge 
have lower exponent of $N$ than the 
the case of (\ref{Db}). 
For the convenience of the next subsection,
we estimate the behavior of  
${\cal D}\Psi \Psi^{\dagger} c_{\alpha} \Psi |initial \rangle$ 
near the boundary.
As a result of ${\cal D}\Psi =0$, the order estimation of $\Psi$ 
near the boundary 
give us 
\begin{eqnarray}
\Psi \sim \left(
\begin{array}{c}
\frac{1}{\sqrt{N}} \\  \frac{1}{\sqrt{N}} \\
 1
\end{array}
\right) \  
\begin{array}{c}
\}k \\  \}k \\
 \}n
\end{array} \ .
\end{eqnarray}
Using this and (\ref{ppd}), the non-vanishing leading term of
${\cal D}\Psi \Psi^{\dagger} c_{\alpha}\Psi $ for a some 
state $|N_1,N_2\rangle $ near boundary $N_1+N_2=N$ is  
\begin{eqnarray}
\label{order0}
{\cal D}&& \!\!\!\!\!\!\!\!   
\Psi \Psi^{\dagger} c_{\alpha}|N_1, N_2 \rangle 
1_{[2k+n]}\langle N_1, N_2| \Psi | initial \rangle \sim
 \label{intermediate} \\  
 &&
\left( 
\begin{array}{c}
\{ B_2(I c_2 \frac{1}{N}-J^{\dagger}c^{\dagger}_1\frac{1}{N})c_{\alpha} 
+B_1(I c_1 \frac{1}{N}+J^{\dagger}c^{\dagger}_2\frac{1}{N})c_{\alpha} 
\} |N_1, N_2 \rangle  \\
\{ -B_1^{\dagger}(I c_2 \frac{1}{N}-J^{\dagger}c^{\dagger}_1\frac{1}{N})
c_{\alpha} 
+B_2^{\dagger}
(I c_1 \frac{1}{N}+J^{\dagger}c^{\dagger}_2\frac{1}{N})c_{\alpha} 
\} |N_1, N_2 \rangle 
\end{array}
\right).
\nonumber 
\end{eqnarray}
Leading terms cancel each other, then only next leading terms 
remain in (\ref{order0}). 
Therefore we can see this expression is not similar to Eq.(\ref{Db}) at all.
The terms like this appear in the integral of the 
Pontrjagin class in the next subsection, 
but it is shown by order estimation of $N$ that 
such terms do not contribute to the integral. \\

For later computation, it is useful to emphasize one fact 
that for arbitrary operand state in $\underbar{{\it D}}$ 
we do not have to distinguish the projector 
$\Psi \Psi^{\dagger}$ from 
$\sum \Psi |n_1, n_2 \rangle \langle n_1,n_2 | \Psi^{\dagger}$
whose region of the summation is some set of intermediate states 
like $\bar{I}$ or $\bar{D}$.
This fact is obtained from the Remark\ref{asymptotic} and the definition
of the boundary. The Remark\ref{asymptotic} 
shows that the $v_0$ have almost no element near the boundary for large $N$, 
then $\Psi$ is an isometry from finite domain $\underbar{{\it D}}$
to $\bar{D}$, and $| \langle n,m | \Psi | n' , m' \rangle |$ is almost zero
if $(n,m) \not{ \!\! \in} \bar{D}$ and $(n',m') \in \underbar{{\it D}}$. 
This is why we do not have to distinguish $\Psi \Psi^{\dagger}$ from 
$\sum \Psi |n_1,n_2 \rangle \langle n_1,n_2 | \Psi^{\dagger}$.\\

At the last of this subsection, 
we emphasize that the result of the calculation is 
independent from the choice of boundary.
We chose artificial boundary in this section
to make calculation be simple.
But, if we perform the same calculation by 
commutative fields with 
using the star product (see for example \cite{Connes1}), 
it is easy to understand that
choosing boundary do not change the result.
So far as the instanton charge is well-defined 
by converge series, the result is identified 
with the result using the star product calculation.
Indeed in the next subsection, essentially 
we use only the fact that the boundary states have 
large eigenvalue of total number operator $N$
and the rank of the boundary states (number of 
the Fock state that construct the boundary)
is $O \!(N)$. (But the notation and calculation 
change into simple one 
as a result of the definition (\ref{domain}).)
So, all we have to do is to make sure that the series 
is converge and identified  with the instaton number
that appear in ADHM construction as a dimension of 
vector space.

\subsection{Direct calculation of Instanton charge }
Let us carry out the calculation of the instanton charge
with only primitive methods.
Using boundary, $D_{\alpha}$ and (\ref{EQ:Q}) the instanton charge
(the integral of the first Pontrjagin class)
 is written as 
\begin{eqnarray}
 Q&=&\lim_{N \rightarrow \infty} Q_N \\
 Q_N&=&-\mbox{Tr}_{N}1+\frac{\zeta^2}{2}\mbox{Tr}_{N}
  \left\{
   [D_1,D_{\bar{1}}][D_2,D_{\bar{2}}] +[D_2,D_{\bar{2}}][D_1,D_{\bar{1}}]
        \right. \label{q-1}\\
 &&\left.\hspace{2.5cm}-[D_1,D_{\bar{2}}][D_2,D_{\bar{1}}]
    -[D_2,D_{\bar{1}}][D_1,D_{\bar{2}}] \right\}\nonumber\\
 &=&-\mbox{Tr}_{N}1+\frac{\zeta^2}{2}\mbox{Tr}_{N}
  \left\{ [D_{\bar{2}},\ D_2 D_{\bar{1}} D_1-D_1 D_{\bar{1}}D_2]
   +[D_{\bar{1}},\ D_1 D_{\bar{2}} D_2-D_2 D_{\bar{2}}D_1] \right\}.
   \label{q-1a}\nonumber
\end{eqnarray}
The boundary that we use is defined by (\ref{Db}) in the previous 
subsection.\\

Let us see the concrete form of these terms.
Note that ${\cal D}\Psi \neq 0$ because 
our calculation is done with
boundary, as we studied in the previous subsection.
The surviving terms  that contain ${\cal D}\Psi$ 
proportional terms exist on the boundary.
So we can use the Eqs.(\ref{Db}),(\ref{order0}) and so on, for these 
surviving terms.
Using (\ref{ppd}), (\ref{ppp1}-\ref{ppp6}) and 
commutation relations, we get following 
explicit expression. 
\\
{\large{ $D_2 D_{\bar{1}} D_1$ : }} 
\begin{eqnarray}
&\zeta^{\frac{3}{2}} & \!\!\!\!\! D_2 D_{\bar{1}} D_1 = \nonumber \\
&-&\Psi^{\dagger} c_2 n_1 \Psi    
\nonumber \\
&+&\Psi^{\dagger} {\cal D}^{\dagger} c_2 
\frac{1}{{\cal D}{\cal D}^{\dagger}}n_1 {\cal D} \Psi   
\nonumber \\
&+&\Psi^{\dagger} \left(
\begin{array}{ccc}
0&0&0 \\
-\zeta \frac{1}{\square} c_1 & 0& 0 \\
0&0&0
\end{array}
\right) \Psi                           
+   \Psi^{\dagger} \left(
\begin{array}{ccc}
0&0&0 \\
0& \zeta c_2 \frac{1}{\square}  & 0 \\
0&0&0
\end{array}
\right) \Psi                              
\nonumber \\                               
&+&\Psi^{\dagger} \tau^{\dagger} c_2 \frac{1}{\square}
(0, \sqrt{\zeta} c_1^{\dagger} , 0 ) \Psi   
+\Psi^{\dagger} \sigma c_2 \frac{1}{\square}
(\sqrt{\zeta} c_1, 0, 0 ) \Psi            
+\Psi^{\dagger} 
\left(
\begin{array}{c}
0 \\ -\sqrt{\zeta} \\ 0 
\end{array}
\right) \frac{1}{\square} n_1 \sigma^{\dagger} \Psi 
\nonumber \\
&+&\Psi^{\dagger} \left(
\begin{array}{c}
0 \\ \sqrt{\zeta} \\ 0 
\end{array} \right)
c_2 \frac{1}{\square} c_1 \tau \Psi 
\nonumber \\
&-&\Psi^{\dagger} {\cal D}^{\dagger} c_2 \frac{1}{{\cal D}{\cal D}^{\dagger}}
{\cal D}   \left(
\begin{array}{c}
0 \\ \sqrt{\zeta} \\ 0 
\end{array} \right) \frac{1}{\square}
c_1 \tau \Psi       \ .             
\label{d211}
\end{eqnarray}
 

{\large $D_1 D_{\bar{1}} D_2 $}:
\begin{eqnarray}
&\zeta^{\frac{3}{2}} & \!\!\!\!\! D_1 D_{\bar{1}} D_2 = \nonumber \\
&-&\Psi^{\dagger} c_1 c_1^{\dagger} c_2  \Psi    
\nonumber \\
&+&\Psi^{\dagger} {\cal D}^{\dagger}(n_1 +1) 
\frac{1}{{\cal D}{\cal D}^{\dagger}} c_2 {\cal D} \Psi   
\nonumber \\
&+&\Psi^{\dagger} \left(
\begin{array}{ccc}
\zeta \frac{1}{\square} c_2 & 0& 0 \\
0&0&0 \\
0&0&0
\end{array}
\right) \Psi                           
+   \Psi^{\dagger} \left(
\begin{array}{ccc}
0&0&0 \\
\zeta c_1 \frac{1}{\square} &0  & 0 \\
0&0&0
\end{array}
\right) \Psi                              
\nonumber \\                               
&+&\Psi^{\dagger} \tau^{\dagger} (n_1+1) \frac{1}{\square}
(\sqrt{\zeta} , 0 , 0) \Psi               
+\Psi^{\dagger} 
\left(
\begin{array}{c}
0 \\ \sqrt{\zeta} c_1 \\ 0 
\end{array}
\right) \frac{1}{\square} c_2 \tau \Psi 
\nonumber \\
&+&\Psi^{\dagger} \left(
\begin{array}{c}
 \sqrt{\zeta} c_1^{\dagger} \\ 0 \\ 0 
\end{array} \right)
 \frac{1}{\square} c_2 \sigma^{\dagger} \Psi 
 +\Psi^{\dagger} \sigma c_1 \frac{1}{\square} 
 (\sqrt{\zeta} c_2 , 0 , 0 )                 
\nonumber \\
&-&\Psi^{\dagger} \sigma c_1 \frac{1}{\square}
(\sqrt{\zeta} , 0 , 0 )
{\cal D}^{\dagger} \frac{1}{{\cal D}{\cal D}^{\dagger}}
c_2 {\cal D} \Psi      \ .              
\label{d112}
\end{eqnarray}
Note that the matrices sandwiched between $\Psi^{\dagger}$ and $\Psi$
are $(k+k+n)\times (k+k+n)$ matrices.
Here we ignore low order terms of $N$ that 
do not contribute to the instanton charge.
For example, 
\begin{eqnarray}
\Psi^{\dagger} {\cal D}^{\dagger} c_2 
({\cal D}{\cal D}^{\dagger})^{-1} {\cal D}
\left(
\begin{array}{ccc}
0&0&0\\
0& \zeta \frac{1}{\square} &0 \\
0&0&0
\end{array}
\right) \Psi =
\Psi^{\dagger}
\left. \left(
\begin{array}{ccc}
0&o(\frac{N_2}{N})&0\\
0& O(\frac{N_2}{N})&0 \\
0&o(\frac{N_2}{N})&0
\end{array}
\right)
\Psi \right|_{boundary}
\end{eqnarray}
is removed from (\ref{d211}).
Because its contribution to the instanton charge is
\begin{eqnarray}
\lim_{N \rightarrow \infty}&& \!\!\!\!\! \!\!\!\!\!
Tr_N \Big[
 \Psi^{\dagger} c_2 \Psi \ , \ 
 \Psi^{\dagger}
\left(
\begin{array}{ccc}
0&o(\frac{\sqrt{N_2}}{N})&0\\
0& O(\frac{\sqrt{N_2}}{N})&0 \\
0&o(\frac{\sqrt{N_2}}{N})&0
\end{array}
\right)
\Psi 
\Big]
\sim 
\lim_{N \rightarrow \infty}
\sum_{boundary} O\left( \frac{1}{N^{\frac{3}{2}}} \right) 
\nonumber \\
&\sim & \lim_{N \rightarrow \infty}
N\ O\!\left( N^{-\frac{3}{2}} \right) =0,
\end{eqnarray}
where $\sum_{boundary}$ denotes the trace over the 
boundary constructed out of $N$ Fock states.
Explicit expression (\ref{ppp1}-\ref{ppp6}) is used here.\\

Using (\ref{d211}) and (\ref{d112}),
\begin{eqnarray}
Q_N &=& -Tr_N 1 +\frac{\zeta^{\frac{3}{2}}}{2}
\left\{ Tr_N [ \Psi^{\dagger} c_2^{\dagger} \Psi \ , \ 
D_2 D_{\bar{1}} D_1-D_1 D_{\bar{1}} D_2 ] +(1\iff 2) \right\}
\nonumber \\
&=& -Tr_N 1 + \frac{1}{2}
\left\{ Tr_N ( A_1 + A_2 +A_3 +A_4)  +(1\iff 2) \right\} ,
\end{eqnarray}
where $(1\iff 2)$ is defined by 
$f(z_1 , z_2 ) -(1\iff 2) \equiv f(z_1 , z_2 )-f(z_2 , z_1 )$,
and $A_1,A_2,A_3$ and $A_4$ are defined by
\begin{eqnarray}
A_1&=&  [ \Psi^{\dagger} c_2^{\dagger} \Psi \ ,\ 
-\Psi^{\dagger} c_2 n_1 \Psi 
+\Psi^{\dagger} c_1 c_1^{\dagger} c_2  \Psi ]
= [ \Psi^{\dagger} c_2^{\dagger} \Psi \ ,\ 
\Psi^{\dagger} c_2 \Psi ]  
\label{A1} \\                     
A_2&=&  \left[ \Psi^{\dagger} c_2^{\dagger} \Psi \ ,\ 
\Psi^{\dagger} \left(
\begin{array}{ccc}
0&0&0 \\
-\zeta \frac{1}{\square} c_1 & 0& 0 \\
0&0&0
\end{array}
\right) \Psi                           
+   \Psi^{\dagger} \left(
\begin{array}{ccc}
0&0&0 \\
0& \zeta c_2 \frac{1}{\square}  & 0 \\
0&0&0
\end{array}
\right) \Psi    \right]              
\nonumber \\
&&- \left[ \Psi^{\dagger} c_2^{\dagger} \Psi \ ,\ 
\Psi^{\dagger} \left(
\begin{array}{ccc}
\zeta \frac{1}{\square} c_2 & 0& 0 \\
0&0&0 \\
0&0&0
\end{array}
\right) \Psi                           
+   \Psi^{\dagger} \left(
\begin{array}{ccc}
0&0&0 \\
\zeta c_1 \frac{1}{\square} &0  & 0 \\
0&0&0
\end{array}
\right) \Psi \right]                               
\nonumber \\
&=& 
 \left[ \Psi^{\dagger} c_2^{\dagger} \Psi \ ,\
\Psi^{\dagger} \left(
\begin{array}{ccc}
-\zeta \frac{1}{\square} c_2 & 0& 0 \\
-\zeta c_1 \frac{2}{\square} & \zeta c_2 \frac{1}{\square}  & 0 \\
0&0&0
\end{array}
\right) \Psi \right]   
\label{A-2} \\              
A_3&=&  \left[ \Psi^{\dagger} c_2^{\dagger} \Psi \ ,\
\Psi^{\dagger} {\cal D}^{\dagger} c_2 
\frac{1}{{\cal D}{\cal D}^{\dagger}}n_1 {\cal D} \Psi   
-\Psi^{\dagger} {\cal D}^{\dagger} c_2 \frac{1}{{\cal D}{\cal D}^{\dagger}}
{\cal D}   \left(
\begin{array}{c}
0 \\ \sqrt{\zeta} \\ 0 
\end{array} \right) \frac{1}{\square}
c_1 \tau \Psi                    
\right]      \label{A-3}  \\
&-& \left[ \Psi^{\dagger} c_2^{\dagger} \Psi \ ,\
\Psi^{\dagger} {\cal D}^{\dagger}(n_1 +1) 
\frac{1}{{\cal D}{\cal D}^{\dagger}} c_2 {\cal D} \Psi   
-\Psi^{\dagger} \sigma c_1 \frac{1}{\square}
(\sqrt{\zeta} , 0 , 0 )
{\cal D}^{\dagger} \frac{1}{{\cal D}{\cal D}^{\dagger}}
c_2 {\cal D} \Psi  \right]           \nonumber       
\end{eqnarray}
\begin{eqnarray}
A_4&=& \left[ \Psi^{\dagger} c_2^{\dagger} \Psi \ ,\
\Psi^{\dagger} \tau^{\dagger} c_2 \frac{1}{\square}
(0, \sqrt{\zeta} c_1^{\dagger} , 0 ) \Psi   
+\Psi^{\dagger} \sigma c_2 \frac{1}{\square}
(\sqrt{\zeta} c_1, 0, 0 ) \Psi            
\right. \nonumber \\
&& \ \ \ \ \ \ \ \ \ \ \ \ \ \ \ \ 
\left. +\Psi^{\dagger} 
\left(
\begin{array}{c}
0 \\ -\sqrt{\zeta} \\ 0 
\end{array}
\right) \frac{1}{\square} n_1 \sigma^{\dagger} \Psi 
+\Psi^{\dagger} \left(
\begin{array}{c}
0 \\ \sqrt{\zeta} \\ 0 
\end{array} \right)
c_2 \frac{1}{\square} c_1 \tau \Psi 
\right]
\nonumber \\
&&  - \left[ \Psi^{\dagger} c_2^{\dagger} \Psi \ ,\
\Psi^{\dagger} \tau^{\dagger} (n_1+1) \frac{1}{\square}
(\sqrt{\zeta} , 0 , 0) \Psi               
+\Psi^{\dagger} 
\left(
\begin{array}{c}
0 \\ \sqrt{\zeta} c_1 \\ 0 
\end{array}
\right) \frac{1}{\square} c_2 \tau \Psi 
\right. \nonumber \\
&& \ \ \ \ \ \ \ \ \ \ \ \ \ \ \ \ 
\left. + \Psi^{\dagger} \left(
\begin{array}{c}
 \sqrt{\zeta} c_1^{\dagger} \\ 0 \\ 0 
\end{array} \right)
 \frac{1}{\square} c_2 \sigma^{\dagger} \Psi 
 +\Psi^{\dagger} \sigma c_1 \frac{1}{\square} 
 (\sqrt{\zeta} c_2 , 0 , 0 )                 
\right] .
\end{eqnarray}

\begin{figure}[t]
  \begin{center}
   \scalebox{0.3}{\includegraphics{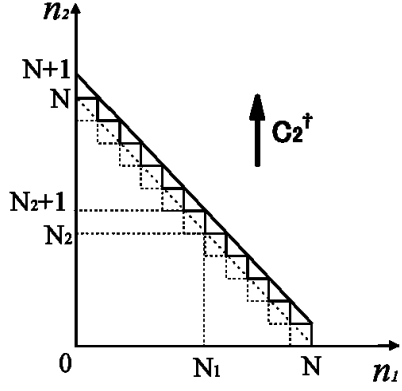}}
  \end{center}
 \caption{$ c_2^{\dagger} \bar{D}$ : $ c_2^{\dagger} \bar{D}$ is 
defined by $\bar{D}$ shifted by $c_2^{\dagger}$.
 Trace over its boundary contribute to the instanton charge.}
 \label{Graph2}
\end{figure}

$A_3$ has intermidiate boundary whose states are operated by
${\cal D}$ or $\tau$ or $\sigma$, like (\ref{ex1}). 
From (\ref{intermediate}),
the trace of $A_3$ is estimated as 
\begin{eqnarray}
\lim_{N \rightarrow \infty}Tr_N A_3 \le 
 C \lim_{N \rightarrow \infty}
N\ O\!\left( N^{-\frac{3}{2}} \right) =0,
\label{A_3}
\end{eqnarray}
where $C$ is some constant.\\
On the other hand, $A_4$ is constructed from the 
commutation relation of the terms like
$\Psi^{\dagger} {\cal D}^{\dagger} \cdots \Psi$ or 
$\Psi^{\dagger}  \cdots {\cal D} \Psi$. 
The trace of such terms is 
\begin{eqnarray}
Tr_N&& \!\!\!\!\!\!\!
 \left[ \Psi^{\dagger} c_2^{\dagger} \Psi \ ,\
\Psi^{\dagger} {\cal D}^{\dagger} \cdots \Psi \right]\\
&=& tr \!\!\!\!\! \!\!
\sum_{{\bf all} \ |initial \rangle} \left\{
\langle initial |  \Psi^{\dagger} c_2^{\dagger} \Psi
\Psi^{\dagger} 
\sum_{{\bf all} \ {\bf int.\!\!\! -b}} 
|{\bf int.\!\!\! -b} \rangle \langle {\bf int.\!\!\! -b} |
 {\cal D}^{\dagger}_b \cdots \Psi | initial \rangle 
 \right. \nonumber \\ &&
\left. -\langle initial |
 \Psi^{\dagger}
 \sum_{N_1+N_2=N} |N_1 , N_2 \rangle \langle N_1 , N_2 |
  {\cal D}^{\dagger} \cdots \Psi   
\Psi^{\dagger} c_2^{\dagger} \Psi | initial \rangle \right\}  
 \nonumber \\
&=&
N \ O\!\left( N^{-\frac{3}{2}} \right)
-Tr_N \left\{
 \Psi^{\dagger}\!\!\!\!
 \sum_{N_1+N_2=N} |N_1 , N_2 \rangle \langle N_1 , N_2 |
  {\cal D}_b^{\dagger} \cdots \Psi   
\Psi^{\dagger} c_2^{\dagger} \Psi \right\}. 
\end{eqnarray}
Therefore trace of $A_4$ is 
\begin{eqnarray}
Tr_N A_4  &=& O\!\left( N^{-\frac{1}{2}} \right)+  \\ 
tr \!\!\!\!\! \!\!
\sum_{{\bf all} \ |initial \rangle}  &&\!\!\!\! \!\!\!\!\! \!\!\!\!
\!\!\! \!\!\!\!\ \!\!\! \!\!\!\!\ 
\langle initial | 
\left\{
\Psi^{\dagger} \!\!\!\!\!\!
\sum_{N_1+N_2=N}\!\!\!\! \!\!
|N_1 , N_2 \rangle \langle N_1 , N_2 |
\left( \begin{array}{ccc}
\zeta c_2 \frac{N_1}{N} & -\zeta c_2{}^2 \frac{1}{N} c_1^{\dagger} &0\\ 
\zeta \frac{N_1}{N} c_1 & -\zeta c_2 \frac{N_1}{N} c_1^{\dagger} &0\\ 
0&0&0
\end{array}
\right) \Psi \Psi^{\dagger} c_2^{\dagger} \Psi
 \right. 
\nonumber \\
-  \Psi^{\dagger} 
c_2^{\dagger} 
 \!\!\!\! \! &\Psi \Psi^{\dagger} & \!\!\!\! \! \left.
\left( \begin{array}{ccc}
0 & 0 &-\sqrt{\zeta} c_1^{\dagger} \frac{1}{N} c_2 J^{\dagger} \\ 
0 & 0 & -\sqrt{\zeta} \frac{N_1}{N} J^{\dagger}\\ 
0&0&0
\end{array}
\right) \!\!\! \sum_{N_1+N_2=N}\!\!\!\! \!
|N_1 , N_2 \rangle \langle N_1 , N_2 | \Psi
 \right\}| initial \rangle . \nonumber
\end{eqnarray}
Near the boundary ($N_1+N_2=N$) we can use concrete form
of $\Psi \Psi^{\dagger}$ (\ref{ppd}), then 
it is shown that 
the last two terms vanish. After all, 
\begin{eqnarray}
\lim_{N \rightarrow \infty} 
Tr_N A_4  = \lim_{N \rightarrow \infty}
O\!\left( N^{-\frac{1}{2}} \right) =0.
\label{A_4}
\end{eqnarray}
From Eqs.(\ref{A_3}) and (\ref{A_4}), we can understand
the survived terms are obtained from $A_1$ and $A_2$.
In the previous discussion about $A_3$ and $A_4$, 
${\cal D}\Psi$ appeared and it made the trace 
operation be a summation over the boundary.
On the other hand, $A_1$ and $A_2$ do not have 
such terms, then we have to calculate $A_1$ and $A_2$ by  
another method. It is Stokes' like theorem.
For some operator 
$\Psi^{\dagger} O \Psi = \sum O_{l_1, p_1,l_2,p_2}
\Psi^{\dagger}|l_1, p_1 \rangle \langle l_2,p_2 |\Psi $, 
commutation relation with $D_{\bar{2}}$ is given as follows,
\begin{eqnarray}
Tr_N [\Psi^{\dagger} c_2^{\dagger} \Psi &,& 
\Psi^{\dagger} O \Psi ] 
\nonumber \\
= \!\!\!\!\!\!\!\!\!\!
\sum_{\scriptsize
\begin{array}{c}
n_1,m_1\in D \\n_2,m_2\in I
\end{array}
 } \!\!\!\!\!
\sum_{\scriptsize
\begin{array}{c}
l_3,p_3 \in \bar{D} \\ 
l_2,p_2 \in \bar{I}
\end{array}
}&& \!\!\!\!\!\!\!\!\!\!
\sum_{l_1,p_1 \in \bar{D}}
 \Psi^{\dagger}_{n_1,m_1,l_1,p_1}
\sqrt{p_1} \Psi_{l_1,p_1-1,n_2,m_2}
\Psi^{\dagger}_{n_2,m_2,l_2,p_2}
O_{l_2,p_2,l_3,p_3}
\Psi_{l_3,p_3,n_1,m_1} 
\nonumber \\ 
- \!\!\!\!\!\!\!\!\!\!
\sum_{\scriptsize
\begin{array}{c}
n_1,m_1 \in D \\ 
n_2,m_2 \in I
\end{array}}\!\!\!\!
\sum_{\scriptsize
\begin{array}{c}
l_3,p_3 \in c_2^{\dagger}\bar{D}\\
l_2,p_2 \in c_2^{\dagger}\bar{I}
\end{array}} && \!\!\!\!\!\!\!\!\!\!\!
\sum_{l_1,p_1 \in \bar{D}}
\Psi^{\dagger}_{n_1,m_1,l_1,p_1}
O_{l_1,p_1,l_2,p_2}
\Psi_{l_2,p_2,n_2,m_2}
 \Psi^{\dagger}_{n_2,m_2,l_3,p_3}
\sqrt{p_3} \Psi_{l_3,p_3-1,n_1,m_1}
\nonumber \\   
=\!\!\!\!\!\!\!
\sum_{\scriptsize
\begin{array}{c}
l_1,p_1\in \bar{D} \\
 l_3,p_3\in \bar{D}
\end{array}
 } && \!\!\!\!\!\!\!\!\!\!
\sum_{l_2,p_2 \in \bar{D}}
 (\Psi\Psi^{\dagger})_{l_3,p_3,l_1,p_1}
\sqrt{p_1} 
(\Psi\Psi^{\dagger})_{l_1,p_1-1,l_2,p_2}
O_{l_2,p_2,l_3,p_3}
\nonumber  \\ 
-\!\!\!\!\!\!\!
\sum_{\scriptsize
\begin{array}{c}
l_2,p_2\in c_2^{\dagger}\bar{D} \\
 l_3,p_3\in c_2^{\dagger}\bar{D}
\end{array}
 } && \!\!\!\!\!\!\!\!\!\!
\sum_{l_1,p_1 \in \bar{D}}
 (\Psi\Psi^{\dagger})_{l_2,p_2,l_3,p_3}
\sqrt{p_3} 
(\Psi\Psi^{\dagger})_{l_3,p_3-1,l_1,p_1}
O_{l_1,p_1,l_2,p_2} ,
\end{eqnarray}
where $ c_2^{\dagger} \bar{D} $ $ (c_2^{\dagger} \bar{I}) $
 is domain defined by $\bar{D}$ $(\bar{I})$ shifted
by $c_2^{\dagger}$ (see Fig.\ref{Graph2}).
Here we omit the indices of 
$2k+n$ dimension vector for simplicity. The 
domain of the trace operation 
$\bar{D}$ is universal among each 
component of $2k+n$ dimension vector
by the definition of the boundary
in the previous subsection (\ref{domain}) and (\ref{trace}).
On the other hand, for intermediate state
there is no universal domain,
i.e. the integral domain of intermediate states 
is determined case by case. So the 
$I$ and $\bar{I}$ are only symbolical
character that means domain of the intermediate states.\\
Only the terms that are on the boundary
shifted by $c_2^{\dagger}$ are remained because they have no partner to 
cancel out. 

Then the remaining terms are written as
\begin{eqnarray}
-\!\!\!\!\!
\sum_{\scriptsize
\begin{array}{c}
l_2,p_2\in \bar{D} \\
 l_3,p_3\in c_2^{\dagger}\bar{D}
\end{array}
 } && \!\!\!\!\!\!\!\!\!\!
\sum_{N_1+N_2=N}\!\!\!\!\!\!
\sqrt{N_2+1} 
 (\Psi\Psi^{\dagger})_{N_1,N_2,l_2,p_2}
O_{l_2,p_2,l_3,p_3}
(\Psi\Psi^{\dagger})_{l_3,p_3-1,N_1,N_2+1}
\label{D3(1)}  \\ 
-\!\!\!\!\!
\sum_{\scriptsize
\begin{array}{c}
l_1,p_1\in \bar{D} \\
 l_3,p_3\in \bar{D}
\end{array}
 } && \!\!\!\!\!\!\!\!\!\!
\sum_{N_1+N_2=N}
 (\Psi\Psi^{\dagger})_{N_1,N_2+1,l_3,p_3}
 \sqrt{p_3} 
 (\Psi\Psi^{\dagger})_{l_3,p_3-1,l_1,p_1}
O_{l_1,p_1,N_1,N_2+1}.
\label{D3(2)}  
\end{eqnarray}

Using this result, let us investigate $A_2$ terms here. 
To adapt (\ref{D3(1)}) and (\ref{D3(2)}) 
for $A_2$,
$O$ should be replaced by
\begin{eqnarray}
\left(
\begin{array}{ccc}
-\zeta \frac{1}{\square} c_2 & 0& 0 \\
-\zeta c_1 \frac{2}{\square} & \zeta c_2 \frac{1}{\square}  & 0 \\
0&0&0
\end{array}
\right). \label{O}
\end{eqnarray}
It is better for simplicity, that (\ref{D3(1)}) and (\ref{D3(2)}) are 
evaluated separately.
As a result of the definition of the boundary,
only non-vanishing term in (\ref{D3(1)}) 
is given as 
\begin{eqnarray}
-Tr_{\scriptsize
boundary
 } &&\left\{ c_2^{\dagger} 
\left(
\begin{array}{ccc}
\cdots& \cdots & \vdots \\
\cdots& \cdots & \vdots \\
I^{\dagger} \frac{1}{\sqrt{\zeta} N} c_2^{\dagger}
&I^{\dagger} \frac{1}{\sqrt{\zeta} N} c_1^{\dagger}&\ddots
\end{array}
\right)             
\left(
\begin{array}{ccc}
-\zeta \frac{1}{\square} c_2 & 0& 0 \\
-\zeta c_1 \frac{2}{\square} & \zeta c_2 \frac{1}{\square}  & 0 \\
0&0&0
\end{array}
\right)     \right.     \nonumber \\  
&& \left. \times \left(
\begin{array}{ccc}
\cdots& \cdots & c_2 \frac{1}{\sqrt{\zeta} N} I\\
\cdots& \cdots & c_1 \frac{1}{\sqrt{\zeta} N} I\\
\cdots& \cdots &\ddots
\end{array}
\right) \right\} \nonumber \\
=&& \frac{1}{\zeta} tr \sum_{boundary} I^{\dagger} \frac{N_2}{N} I
\label{D3(1')}  
\end{eqnarray}
Here $Tr_{boundary}$ denotes a summation over the 
boundary of $c_2^{\dagger} \bar{D}$ and we ignore 
some components that do not contribute to the integral
in the large $N$ limit. $tr$ represent trace operation for the all indices
without Fock space indices i.e. $\makebox{tr}_{U(N)}$. \\

Let us evaluate (\ref{D3(2)}) in $A_2$.
As a result of the boundary definition, 
we can conclude that 
$\langle l_3,p_3-1| c_2^{\dagger} |N_1,N_2 \rangle=0$
if $( l_3,p_3-1) \in \bar{D}$.
Then (\ref{D3(2)}) whose $O$ is replaced 
by (\ref{O}) is
\begin{eqnarray}
-tr \sum_{N_1+N_2=N}&& 
\left\{ \langle N_1,N_2 | \frac{c_2}{\sqrt{N_2}}
\left(
\begin{array}{ccc}
\cdots& \cdots & - \frac{1}{\sqrt{\zeta} N} c_1^{\dagger}J^{\dagger}\\
\cdots& \cdots & c_2^{\dagger} \frac{1}{\sqrt{\zeta} N}J^{\dagger} \\
I^{\dagger} \frac{1}{\sqrt{\zeta} N} c_2^{\dagger}
&I^{\dagger} \frac{1}{\sqrt{\zeta} N} c_1^{\dagger}&1_{[n]}
\end{array}
\right) \right.      
\nonumber \\
  \times \!\!\!
&c_2^{\dagger}& \!\!\!\!\!
\left.
\left(
\begin{array}{ccc}
\cdots& \cdots & 0\\
\cdots& \cdots & 0 \\
-J \frac{\sqrt{(N_2+1)}}{\sqrt{\zeta} N^2} c_1
&I^{\dagger} \frac{\sqrt{(N_2+1)}}{ \sqrt{\zeta}  N^2} c_1^{\dagger}
+J \frac{\sqrt{ (N_2+1)}}{ \sqrt{\zeta}  N^2} c_2&0
\end{array}
\right) |N_1,N_2 \rangle 
\right\}
\nonumber \\
&=& - \frac{1}{\zeta}\    tr \sum_{boundary} \left( 
J\frac{N_2^2 + N_1 N_2}{N^3} J^{\dagger}
\right), 
\label{D3(2')}
\end{eqnarray}
where we use $tr(IJ)=-tr([B_1,B_2])=0$. From (\ref{D3(1')}) 
and (\ref{D3(2')}), 
\begin{eqnarray}
Tr_N (A_2 +(1 \iff 2))
=Tr_{boundary}\frac{1}{\zeta N} (I^{\dagger} I-JJ^{\dagger})
\nonumber \\
= \frac{1}{\zeta} tr (I^{\dagger} I-JJ^{\dagger}) =2k .
\label{A_2}
\end{eqnarray}
Here we use trace of the real ADHM equation (\ref{EQ:ADHM1}).

Final work we have to do is to evaluate $A_1$.
Using (\ref{pp}) and commutation relations
\begin{eqnarray}
A_1 &=& -{\zeta}[D_{\bar{2}}\ ,\ D_2 ] \\
&=&- 1  \label{-1}\\                                       
    &&-  \Psi^{\dagger} {\cal D}^{\dagger} c_2^{\dagger} 
      \frac{1}{{\cal D}{\cal D}^{\dagger}}c_2 {\cal D} \Psi        
    +  \Psi^{\dagger} {\cal D}^{\dagger} c_2 
      \frac{1}{{\cal D}{\cal D}^{\dagger}}c_2^{\dagger} {\cal D} \Psi 
    \label{2_2'}\\
   &&-   \Psi^{\dagger} \tau^{\dagger} c_2^{\dagger} 
      \frac{1}{\square}(\sqrt{\zeta} , 0 , 0)\Psi  
    -   \Psi^{\dagger} \left(
       \begin{array}{c}
       \sqrt{\zeta} \\ 0 \\ 0
       \end{array}
       \right)
       \frac{1}{\square}c_2 \tau \Psi           
    \label{3}\\   
    &&+\Psi^{\dagger} \sigma c_2 
      \frac{1}{\square}(0, -\sqrt{\zeta} , 0)\Psi 
     +  \Psi^{\dagger} \left(
       \begin{array}{c}
       0\\ -\sqrt{\zeta} \\ 0 
       \end{array}
       \right) \frac{1}{\square}c_2^{\dagger} 
       \sigma^{\dagger} \Psi           
     \label{4}\\
    &-&\!\!\!\! \Psi^{\dagger} \left(
       \begin{array}{c}
       \sqrt{\zeta} \\ 0 \\ 0
       \end{array}
       \right)
       \frac{1}{\square}  (\sqrt{\zeta} , 0 , 0)\Psi 
     +  \Psi^{\dagger} \left(
       \begin{array}{c}
       0\\ \sqrt{\zeta} \\ 0 
       \end{array}
       \right) \frac{1}{\square}   
       (0, \sqrt{\zeta} , 0)\Psi    \label{1}            
\end{eqnarray}

Trace of (\ref{2_2'}) is represented as follows:
\begin{eqnarray}
Tr_N &&\!\!\!\!\!\!\!
 \left\{-  \Psi^{\dagger} {\cal D}_b^{\dagger} c_2^{\dagger} 
      \frac{1}{{\cal D}{\cal D}^{\dagger}}c_2 {\cal D}_b \Psi        
    +  \Psi^{\dagger} {\cal D}_b^{\dagger} c_2 
      \frac{1}{{\cal D}{\cal D}^{\dagger}}c_2^{\dagger} {\cal D}_b \Psi \right\}
    \nonumber \\
  = Tr_N && \!\!\!\!\!\!\!\!
   \left\{   
  \Psi \left(
  \begin{array}{cc}
  -\sqrt{\zeta} c_2 & -B_1 \\
  -\sqrt{\zeta} c_1 &  B_2 \\
  0 & J
  \end{array}
  \right)
  \left( \begin{array}{cc}
  c_2 \frac{1}{\square} c_2^{\dagger} - 
  c_2^{\dagger}\frac{1}{\square}c_2 & 0 \\
  0 & c_2 \frac{1}{\square} c_2^{\dagger} - 
  c_2^{\dagger}\frac{1}{\square}c_2
  \end{array} \right) \right.
   \nonumber \\       
  && \times \left.
  \left( \begin{array}{ccc}
  -\sqrt{\zeta} c_2^{\dagger} & -\sqrt{\zeta} c_1^{\dagger}& 0\\
  -B_1^{\dagger} & B_2^{\dagger} & J^{\dagger}
  \end{array}\right) \Psi
  \right\}. \label{22'}
\end{eqnarray}
After straightforward calculation with using concrete form of
$1/\square$ of (\ref{sq_-1}),
(\ref{22'}) become 
\begin{eqnarray}
Tr_{boundary} &&\!\!\!\!\!\! \left\{
\left( -1 +\frac{2N_2^2 + 2 N_1 N_2}{N^2} \right)
\left( \frac{1}{N}-\frac{1}{\zeta}II^{\dagger}\right)
\right. \nonumber \\ 
&+&\frac{(N_1-N_2)^2}{N^3} 
+\frac{-2N_2(N_2-N_1)^2}{N^4} \nonumber 
\\
&+&  2\frac{N_1 N_2}{ \zeta N^2}
    \left(\frac{1}{N}-\frac{2N_2}{N^2} \right)
     \left(II^{\dagger} - J^{\dagger} J\right)
\nonumber \\     
&+& \left. \frac{1}{\zeta} 
J^{\dagger}\left(\frac{1}{N}-\frac{2N_2}{N^2} \right)J
+ O(N^{-\frac{3}{2}})\right\}. \label{22''}
\end{eqnarray}
This is non-vanishing term in the large $N$ limit,
but, when we add $(1 \iff 2)$ to  (\ref{22''}), 
\begin{eqnarray}
\lim_{N \rightarrow \infty} 
\left\{ \hbox{(\ref{22''})} +(1 \iff 2) \right\} = 0.
\label{22'''}
\end{eqnarray}
After similar calculation, (\ref{3}) and its 
$ (1 \iff 2) $ are obtained:
\begin{eqnarray}
2 Tr_{boundary} \left\{ 
\frac{\frac{1}{\zeta}II^{\dagger}-1}{N}
+\frac{(N_1-N_2)^2}{N^3}
+\frac{2N_1 N_2}{\zeta N^3} (II^{\dagger}-J^{\dagger}J)
\right\}.
\label{3'}
\end{eqnarray}
(\ref{4}) and its 
$ (1 \iff 2) $ are 
\begin{eqnarray}
-\frac{2}{\zeta} Tr_{boundary} \left\{ J^{\dagger} \frac{1}{N} J \right\}
\label{4'}.
\end{eqnarray}
The other terms of $A_1$ are (\ref{-1}) and (\ref{1}),
and we can show that 
they vanish by using the similar transformation from ADHM equations and ADHM data
to Instanton(ASD) equation:
\begin{eqnarray}
\left\{-1-
        \Psi^{\dagger} \left(
       \begin{array}{c}
       \sqrt{\zeta} \\ 0 \\ 0
       \end{array}
       \right)\!\!\!\!\! \right.
       &&\!\!\!\!\!\!\!\left.
       \frac{1}{\square} 
        (\sqrt{\zeta} , 0 , 0)\Psi 
     +  \Psi^{\dagger} \left(
       \begin{array}{c}
       0\\ \sqrt{\zeta} \\ 0 
       \end{array}
       \right) \frac{1}{\square}   
       (0, \sqrt{\zeta} , 0)\Psi \right\} 
       +(1\!\!\! \iff \!\!\! 2) \nonumber \\
  &=&   -\zeta( [D_{\bar{2}},D_2]+ [D_{\bar{1}},D_1] )=-2
\label{1'}
\end{eqnarray}

 From the equations (\ref{22'''}),(\ref{3'}),(\ref{4'})
 and (\ref{1'}),
the contribution from $A_1+(1 \iff 2)$ is given as
\begin{eqnarray}
Tr_N(A_1 +(1 \iff 2)) 
&=&-Tr_{boundary} \frac{4}{N}1_{[k]} + 2Tr_N 1 
\nonumber \\
&=& -4k + 2Tr_N 1,
\label{A_1}
\end{eqnarray}
where we use the relation 
$tr (I^{\dagger} I -JJ^{\dagger})= 2\zeta tr 1_{[k]}$, again.\\

After all, from (\ref{A_1}),(\ref{A_2}),(\ref{A_3})
and (\ref{A_4}), 
\begin{eqnarray}
Q_N= -k + O(N^{-\frac{1}{2}}).\ \ Q=\lim_{N \rightarrow \infty} Q_N =-k
\end{eqnarray}
The proof is completed.

\section{Summary and Discussion} 
We have studied the instanton number of U(n) gauge theory
on the noncommutative ${\bf R}^4$ in this article.
From our observation of the $\Psi \Psi^{\dagger}$, 
it was discovered that
there are zero-mode $v_0$ whose dimension of the Fock space is $k$-dim
(see the Lemma\ref{lemma:zero-mode}).
The origin of the instanton number was understood by the 
zero-mode $v_0$.  Further, its asymptotic behavior was investigated 
and we found that it damp faster than exponential damp.   
So we could introduce the boundary for the integral (trace) 
without no difficulty from the zero-mode.
Using the boundary, we showed that we can define the integral of the 
first Pontrjagin class as a converge series.
After direct calculation, we proved the Theorem\ref{theorem1}.
In this proof, we do not use the Lemma\ref{lemma:zero-mode}.
The calculations of section 4 show that we can understand the origin 
of the instanton number by $tr ( I^{\dagger} I-JJ^{\dagger})/\zeta =2k$,
 instead of $v_0$. Although this is unique character of
ADHM construction on noncommutative space,
this theorem implies that the instanton number is defined as same as 
commutative case. 
Especially, the fact that instanton charge is given  by some integer is 
sign of topological nature.
Our calculation is done for only noncommutative ${\bf R}^4$, so 
we cannot conclude that the instanton number or some other characteristic class
is topological invariant.
But, it is natural to expect that noncommutative manifold inherit 
topological invariant from commutative manifold. 
(There is similar circumstantial evidence about topological field theory case 
\cite{Sako1,Sako2}. From these article, partition function of the cohomological
field theory on noncommutative ${\bf R}^n$ is independent from noncommutative
parameter.)  \\

Here we review the proof of Theorem\ref{theorem1} from the view point of 
the space-time noncommutativity.  
The proof was done with the concrete form of $1/\square $ given by
(\ref{sq_-1}). 
Pay attention for the noncommutative parameter $\zeta$ in (\ref{sq_-1}).
We have considered only the case whose noncommutativity
is given by $\theta^{12} = \theta^{34}= -\zeta$ , that is called ``self-dual".
In the Eq.(\ref{sq_-1}), $1/\square $ is expanded by $(\zeta N)^{-1}$.
If we consider $\theta^{12} \neq \theta^{34}$ case,  
$1/\square $ should be expanded by $(\theta^{12} n_1 + \theta^{34}n_2)^{-1} $
and there is no problem so far as the condition $\theta^{12} \theta^{34} >0$.
In the proof of section 4, there are no obstruction for changing 
the noncommutativity to $\theta^{12} \neq \theta^{34}$ under the condition 
$\theta^{12} \theta^{34} >0$. So the theorem is rewrite as following.
\begin{theorem}[Instanton number]
Suppose U(n) gauge theory on noncommutative ${\bf R^4}$ 
whose noncommutative parameters (\ref{parameter}) obey the condition 
$\theta^{12} \theta^{34} >0$. 
Then
the integral of the first Pontrjagin class is possible to 
be defined by converge series and it is identified with
the dimension $k$ of the vector space in the ADHM construction
and that is called ``instanton number".
\end{theorem}

On the other hand when the noncommutativity is 
given by $\theta^{12} \theta^{34} <0$, that is called anti-self-dual,
$1/\square $ expansion by  $(\theta^{12} n_1 + \theta^{34}n_2)^{-1} $
is not effective
\footnote{ After this preprint(ver.2) appeared, Tian, Zhu and Song propose
\cite{Tian}. In the \cite{Tian}, they calculate instanton charge including
 $\theta^{12} \theta^{34} <0$ case.
They use Corrigan's identity in the calculation 
that is well known way in commutative case.
When we use Corrigan's identity in noncommutative theory, 
we have to pay attention for total divergent terms.
Because cyclic symmetry of the trace operation 
is used in the proof of  Corrigan's identity of commutative theory.
In general, when we use the cyclic symmetry in noncommutative theory,
total divergent terms appear.
For avoiding evaluating the total divergent terms,
primitive (but complex) calculation is done in this paper.
Then the role of noncommutativity in the instanton charge is clarified. }
(Such cases have some special nature of the noncommutative instanton.
For example, though this case contain the non-stable instanton moduli space,
instanton solution exist as non-singular function.  
 Such phenomena is studied in \cite{Furuuchi3, Hamanaka}.) 
Because $(\theta^{12} n_1 + \theta^{34}n_2)$ is  
not only always non-large but also sometime almost zero 
 near the boundary, even if we put arbitral large $N$ boundary.
Therefore our proof in this article do not work in this case.
To define the instanton charge as a converge series and to prove
the theorem for this case, 
we have to change the choice of domain of trace operation and to change 
the way of taking a limit.
We do not do it in this article.\\

It has been clarified that the instanton number of noncommutative 
${\bf R}^4$ has the algebraic origin, but its analyze is not enough.
Because, in our case, instantons do not smooth connect to the commutative
instantons in the commutative limit, $\zeta \rightarrow 0$.
But the result about instanton number is not changed.
We have to understand the reason why the instanton number is the same value
as instanton number of commutative ADHM construction by more geometrical picture.
But there is no good idea for it until now.
Further if such characteristic class has topological nature, 
it is expected that there is some kind of relations with the K-theory,
but it is unclear, too.
These analysis are left for future works.\\

{\bf Acknowledgments } \\
I would like to thank Dr.T.Ishikawa, Dr.S.Kuroki and Dr.E.Sako for  
many supports. Discussions during the YITP workshop``Quantum Field 
Theory 2002" were helpful to complete this work. 


\end{document}